\documentclass[twocolumn,aps,pra,showpacs,amssymb,superscriptaddress,longbibliography]{revtex4-1}

\usepackage{graphicx}
\usepackage{amsmath}
\usepackage{amssymb}
\usepackage{bm}
\usepackage[usenames]{color}
\usepackage{amsfonts}
\usepackage{appendix}
\usepackage{footmisc}
\usepackage{multirow}
\usepackage{subcaption} 

\usepackage{comment}
\usepackage[table]{xcolor}

\usepackage[colorlinks=true , citecolor=blue,urlcolor=blue]{hyperref}

\newcommand{\vect}[1]{\boldsymbol{#1}}

\newcommand{\e}{\textrm{e}}

\newcommand{\rbf}{\mathbf{r}}
\newcommand{\vbf}{\mathbf{v}}

\newcommand{\gbf}{\mathbf{g}}

\newcommand{\abf}{\mathbf{a_S}}
\newcommand{\Ombf}{\mathbf{\Omega_S}}

\newcommand{\kbf}{\mathbf{k}}

\begin{document}
\title{Integration of a high-fidelity model of quantum sensors with a map-matching filter for quantum-enhanced navigation}
\author{Samuel Lellouch}
\email[]{s.lellouch.1@bham.ac.uk}
\affiliation{School of Physics and Astronomy, University of Birmingham, Birmingham, B15 2TT, United Kingdom}
\affiliation{School of Engineering, University of Birmingham, Birmingham, B15 2TT, United Kingdom}
\author{Michael Holynski}
\affiliation{School of Physics and Astronomy, University of Birmingham, Birmingham, B15 2TT, United Kingdom}

\begin{abstract}
Harnessing the potential of quantum sensors to assist in navigation requires enabling their operation in complex, dynamic environments and integrating them within existing navigation systems. 
While cross-couplings from platform dynamics generally degrade quantum measurements in a complex manner, navigation filters would need to be designed to handle such complex quantum sensor data.
In this work, we report on the realization of a high-fidelity model of an atom-interferometry-based gravity gradiometer and demonstrate its integration with a map-matching navigation filter. 
Relying on the ability of our model to simulate the sensor behaviour across various dynamic platform environments, we show that aiding navigation via map matching using quantum gravity gradiometry results in stable trajectories, and highlight the importance of non-Gaussian errors arising from platform dynamics as a key challenge to map-matching navigation. 
We derive requirements for mitigating these errors, such as maintaining sensor tilt below $3.3^o$, to inform future sensor development priorities.
This work demonstrates the value of an end-to-end approach that could support future optimization of the overall navigation system.
Beyond navigation, our atom interferometer modelling framework could be relevant to current research and innovation endeavours with quantum gravimeters, gradiometers and inertial sensors.
\end{abstract}

\date{\today}
\maketitle


\section{Introduction}
\label{sec:intro}

Navigating accurately and reliably in environments where traditional positioning systems fail is a critical challenge in both civilian and military applications. Enabling resilient, continuous and precise tracking could enhance safety and efficiency, while also facilitating autonomous navigation systems capable of responding in real-time without human intervention. Conventional navigation methods, such as position fixing via the Global Navigation Satellite System (GNSS), are vulnerable to interference, spoofing and jamming and require a line of sight to satellites, making them unsuitable for denied environments. 
Inertial Navigation Systems (INS), which estimate position, velocity, and orientation by integrating measurements of acceleration and angular velocity from accelerometers and gyroscopes, are widely used in these scenarios due to their ability to operate independently of external signals. However, their performance degrades over time due to the accumulation of measurement and integration errors, resulting in significant drift~\cite{Jekeli2005}. Mitigating these drifts has led to the development of position-fixing methods, which use external sources to perform periodic corrections and maintain navigational accuracy~\cite{Gleason1995,Jekeli2006}. Alternative navigation via gravity gradient map-matching, where observations of the local gravity gradient at different locations are matched with existing database, is particularly promising as an autonomous, stable and passive position fixing method, as gravity is ubiquitous on Earth and immune to jamming or spoofing~\cite{Wang2016,Wang2017,Wu2017,Phillips2022}. 
Although gravity map-matching has been demonstrated using classical accelerometers~\cite{Wang2016b,Wang2017}, it remains challenging due to the small signals of the gravity variations to detect.

Quantum sensors based on cold-atom interferometry~\cite{Riehle1991,Kasevich1991,Kasevich1992}, which have evidenced unprecedented performance levels in measuring inertial quantities~\cite{Gustavson1997,Canuel2006,Barrett2016,Yankelev2019,Geiger2020,Salducci2024} in laboratories, including gravitational acceleration~\cite{Peters1999,Peters2001,DeAngelis2008,Lellouch2022,Tino2021}, have emerged as promising candidates for alternative navigation. 
By using atoms, which are immune to manufacturing defects, wear and tear, as test masses, they not only bring improvement in accuracy over classical sensors~\cite{Peters1999,Peters2001,Hu2013}, but also more practicality as they do not require as frequent calibration to retain their long-term stability~\cite{Durfee2006,Gillot2014}.
Their development has prompted successful demonstrations of quantum gravimetry in the field including on trucks~\cite{Wang2022}, air~\cite{Bidel2020} and marine~\cite{Bidel2018} platforms.
Additionally, operating them in a gradiometric configuration where a common laser beam is used to extract information from two spatially-separated atom interferometers derives additional benefits through common-mode noise rejection~\cite{McGuirk2002,Sorrentino2014,Biedermann2015}. 
Common-mode suppression of 140 dB has been demonstrated in laboratories~\cite{McGuirk2002} before being recently achieved in practical settings, which, together with 600-fold reduction in tilt susceptibility over a perfect gravimeter, has facilitated successful trials in real-life environments~\cite{Stray2022}. 
While this progress has triggered an intense effort to bring quantum gravity gradiometers to practical use in real application contexts~\cite{Schmidt2011,Menoret2018,Bongs2019,Wu2019,AntoniMicollier2022,Janvier2022}, such systems are still in an early development stage compared to classical sensors~\cite{Jekeli2005,Fang2012}.

In particular in navigation applications, quantum sensors are required to operate at high performance in entirely different conditions than they are in lab-based environments. 
Constraints of size, weight and power (SWAP), and bandwidth are often competing with sensitivity requirements.
Additionally, platform motion and influences from the external environment generally lead to significant noise and systematic effects~\cite{Bidel2020,Barrett2016,Barrett2016b,Saywell2023,Darmagnac2024}, whose impact on sensor accuracy, and subsequently on the overall performance of a full gravity-aided navigation system, is yet to be fully quantified.

This lack of understanding is hindering the rapid deployment of quantum sensors in navigation settings and constitutes a major obstacle to their integration with existing INS.
The challenges of integrating quantum sensors into INS architectures are ubiquitous to a range of quantum-enhanced navigation solutions, from quantum gravity-enhanced navigation such as based on map-matching~\cite{Wang2016,Wang2017,Wu2017,Phillips2022} or gravity compensation~\cite{Welker2013}, to the fusion of classical and quantum inertial capabilities within hybrid inertial sensors~\cite{Wright2022,Tennstedt2023,Wang2023b}.
In all these examples, effective integration critically depends on the ability to achieve real-time data fusion, including handling calibration systematics, noise and errors on quantum measurements~\cite{Wright2022}. As these errors become particularly pronounced on moving platforms, how they propagate through INS architectures can be non-trivial. As a result, access to comprehensive sensor datasets is essential for training and optimizing navigation filters to the specificities of each platform and sensor capabilities.

With trials being scarce and costly, modelling is an approach of choice. While classical sensors benefit from decades of modelling and compensation strategies, equivalent models for quantum sensors are still in early development.
Significant progress has been made to understand the various contributions to the interferometric phase shift~\cite{Peters2001,Barrett2016b}, including understanding errors arising from dynamic conditions~\cite{Darmagnac2024}. However, accurately capturing the full scope of interactions and cross-couplings at the microscopic level remains a significant challenge.
Existing simulations of quantum-enhanced navigation so far have relied on generic synthetic sensor data~\cite{Phillips2022}, allowing to demonstrate effective position fixing methods yet under idealised sensor noise assumptions.
Accurate quantum sensor models representative of the sensor true performance in real navigation conditions would therefore be critical to provide comprehensive, reliable data to train and optimize navigation filters.\\

In this paper, we demonstrate the integration of a high-fidelity, microscopic model of a quantum sensor with a map-matching filter, enabling end-to-end simulation of a complete 2D position-fixing solution. Our cold-atom gravity gradiometer model captures a broad spectrum of factors relevant to navigation, accounting for diverse operational environments, including maritime and aerial platforms, as well as the ability to operate with a full range of sensor specifications. Through our simulations, we identify key challenges that must be addressed to advance the use of quantum sensors in navigation. This work constitutes a first significant step towards the future platform-specific optimization of quantum sensors and navigation filters.\\

This paper is organized as follows. In Sec.~\ref{sec:Model}, we introduce our cold-atom gravity gradiometer modelling framework and its implementation. Section~\ref{sec:GGMeasurements} presents digital simulations of the quantum sensor operating along various platform trajectories, before Sec.~\ref{sec:MM} demonstrates integration of these results into a map-matching algorithm to evaluate the potential of quantum-enhanced gravity gradient map-matching navigation. Section~\ref{sec:discussion} discusses the findings and outlines key challenges for future advancements of the technology.

\section{A high-fidelity model for a cold-atom gravity gradiometer}
\label{sec:Model}

\subsection{Physics of an atom-interferometry based gravity gradiometer}
\label{sec:AIbasis}

\begin{figure*}[!t]
\centering
		\includegraphics[width=18cm]{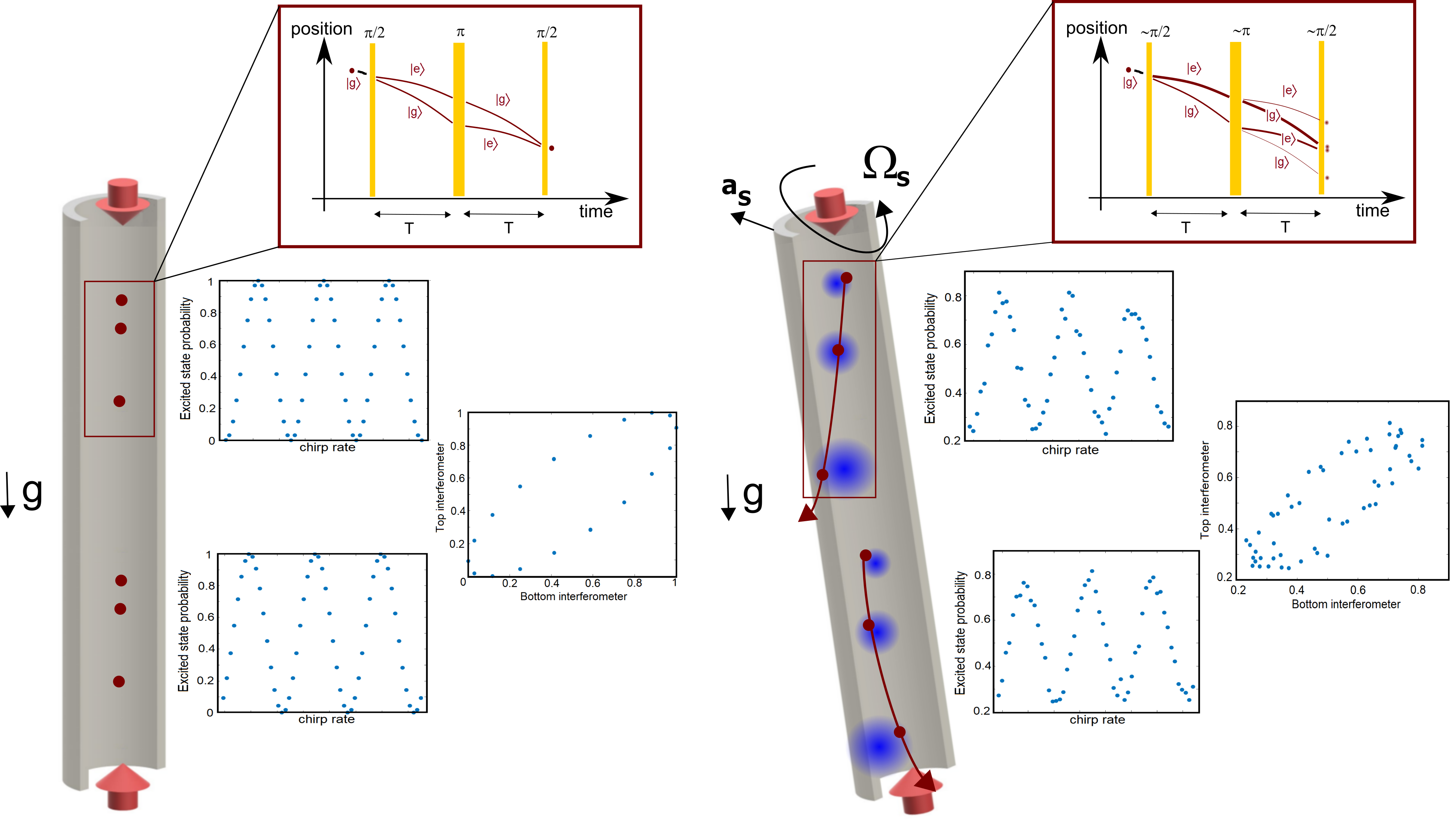}
		\caption{Schematic depiction of the measurement process and mechanisms at play in an atom-interferometry-based gravity gradiometer. Left: In an idealized situation (single-atom, static environment), the beamsplitter and mirror pulses forming the Mach-Zehnder sequence (shown in the red frame) realise perfect population transfers; the fringe contrast is 1 and the differential phase between the two interferometers is imputable to the vertical gravity gradient. Right: In real environments, cross-couplings between atomic cloud inhomogeneities and platform motion create complex individual atomic responses which result in imperfect population transfers; cloud averaging at the detection stage leads to non-trivial phase shifts, contrast loss and noise that couple in with the gravity gradient measurement.
      \label{fig:Fig1}}
\end{figure*}

The quantum sensor considered in this work is a gravity gradiometer based on atom-interferometry. Such sensor is made of two spatially-separated atom interferometers addressed by a common laser beam.
The measurement process is depicted on Fig.~\ref{fig:Fig1}. Within each atom interferometer, previously cooled down atoms are addressed by a sequence of three light pulses while they free-fall in the gravity field. A first $\pi/2-$pulse places the atoms in a quantum superposition of two momentum states, resulting in two wavepackets moving along two separate trajectories. After a time $T$ of free-fall, a $\pi-$pulse is applied that redirects the wavepackets towards each other. The atoms are let to free-fall again for a duration $T$ before a final $\pi/2-$pulse is applied that overlaps the trajectories. Within each interferometer, the atomic state populations display interference fringes as a function of the accumulated phase difference between the two arms of the interferometer. For atoms in free-fall in the gravity field $\mathbf{g}$, this interferometric phase is given by 
$\Phi=\mathbf{g}.\mathbf{\kbf}T^2$ with $\hbar\mathbf{\kbf}$ the momentum splitting of the interferometer~\cite{Kasevich1991}, offering a way to measure the absolute value of the local acceleration due to gravity along the beams axis. Using two vertically-aligned interferometers, the gradiometric configuration then probes the $z-z$ component of the gravity gradient tensor, while conveniently suppressing noise common to both interferometers~\cite{McGuirk2002,Sorrentino2014}. In practice, this measurement is achieved by plotting the two interferometer fringes one versus each other [see Fig.~\ref{fig:Fig1}], producing an elliptical Lissajous plot whose eccentricity and rotation angle give access to the differential phase shift.\\

This simple picture is however highly challenged in complex environments such as those encountered in navigation contexts. 
Platform motion, including tilts, accelerations and rotations, result in the sensor frame being non-inertial, and thereby in the atoms undergoing a much more complex dynamics than a mere vertical free-fall [see Fig.~\ref{fig:Fig1}, right]. 
Inertial effects impact interferometric measurements in multiple ways, leading both to additional phase shifts and to a loss of contrast due to the motion of the atoms in the sensor frame.
Taken separately, the individual contributions of various inertial effects to the interferometric phase are well known at the single-atom level; for instance, a uniform sensor acceleration $\mathbf{a_S}$ produces an accelerational phase shift $\Phi=\mathbf{a_S}.\mathbf{\kbf}T^2$ while a constant rotation $\mathbf{\Omega_S}$ produces a rotational phase shift $\Phi=(2\mathbf{\Omega_S}\times\mathbf{v}).\mathbf{\kbf}T^2$ with $\mathbf{v}$ the atomic velocity~\cite{Geiger2020}.
However, cross-couplings between various inertial effects make the overall phase shift highly non-trivial.
For instance, accelerations modify atomic velocities, thereby impacting the rotational phase shift, while rotations, by modifying the axis of the sensor, not only change the instantaneous projection of acceleration fields on the laser axis but also create, in a gradiometric configuration, differential accelerations between the two interferometers.
More generally, cross-couplings between inertial effects and interferometric pulses give rise to complex mechanisms. By affecting the motion of the atoms in the beam, inertial effects directly impact the fidelity of atom-light pulses, inducing both contrast loss and additional phase shifts [see Appendix~\ref{sec:AppSubsec2}]. Reciprocally, momentum kicks imparted at each pulse on a given interferometric arm couple with inertial forces to induce complex non-inertial trajectories of the two wavepackets, resulting in their imperfect overlap at the final $\pi/2$ pulse, further reducing the contrast.\\

A further level of cross-couplings arise from the inhomogeneous nature of atomic cloud, comprised of typically $\approx10^5$ atoms, with an approximately Gaussian cloud shape. 
The individual position of each atom within the beam, which determines the local intensity it experiences, and its individual velocity, which determines the laser frequency it sees as shifted by the Doppler effect, result in each atom experiencing the $\pi-$ and $\pi/2-$pulses differently. 
Even in static systems, these inter-atom dephasings result in contrast loss and an apparent phase shift at the output of the interferometer~\cite{Szigeti2012,Butts2013,Dunning2014,Lellouch2023}.
In dynamic conditions, inter-atom dephasings introduce cross-coupling effects between the atomic cloud and inertial forces, as each atom inside the cloud experiences not only light fields, but also inertial forces, in a different way.
This results in a highly non-trivial interferometric output after cloud-averaging occurs at the detection stage. In practice, cross-couplings between the external environment and inhomogeneous clouds can severely degrade the accuracy and sensitivity of measurements, possibly resulting in systematic effects blurring out the signal of interest~\cite{Geiger2011,Bidel2018,Bidel2020,Saywell2023}.

\begin{figure*}[!t]
\centering
		\includegraphics[width=18cm]{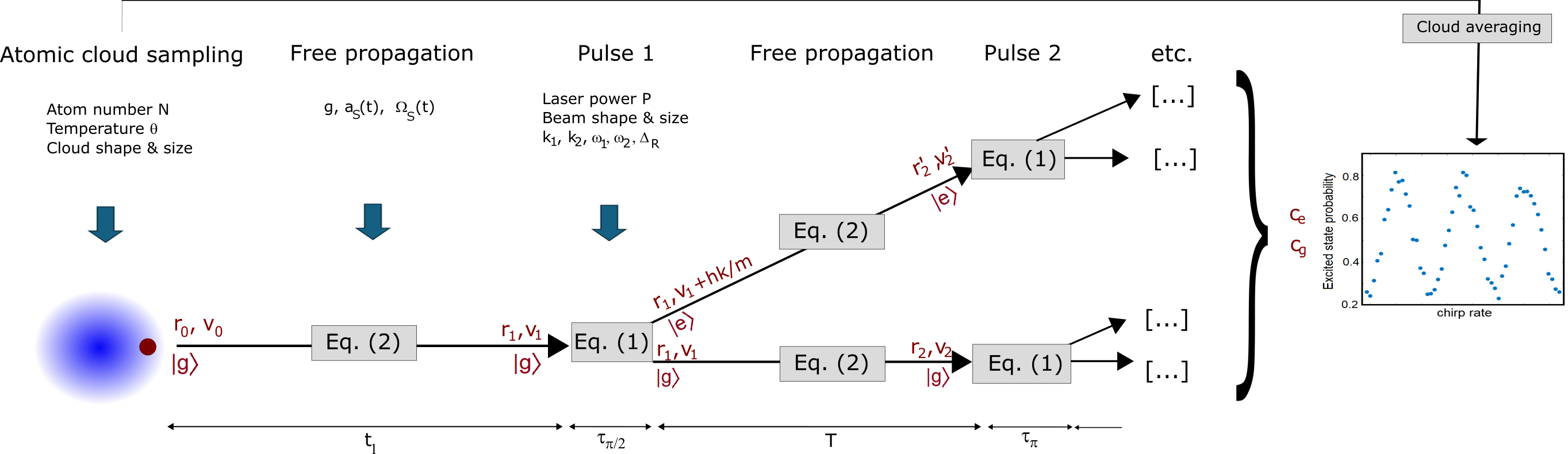}
		\caption{Schematic depiction of the atom interferometer model simulation sequence. Performing a Monte-Carlo sampling of the initial cloud, atomic position, velocity and internal state are propagated in time through alternating periods of pulse interaction and ballistic propagation, along all arms of the interferometer. Final state populations are reconstructed at the end of the sequence and averaged over the initial cloud.
      \label{fig:Fig2}}
\end{figure*}
Cross-coupling effects are notoriously challenging to describe and quantify accurately, due to the complex interplay between internal state evolution, external atomic trajectories and matter-light pulse dynamics.
Traditional atom interferometry models rely on the classical action or phase shift formalism to compute and analyse phase differences between interferometer arms using path integral approaches~\cite{Kasevich1992,Storey1994} or the sensitivity function formalism~\cite{Legouet2008}.
Such models have been applied in dynamic situations, allowing to successfully account for the effects of tilts and rotations~\cite{Darmagnac2024} and to capture inertial phase shifts via integrating atomic kinematics into sensitivity function analysis~\cite{Barrett2016b}. 
However, these approaches typically disregard the effects of pulse infidelities [see Appendix~\ref{sec:AppSubsec2}] and neglect recoil-induced trajectory changes. Additionally, they remain based on a single-atom description that cannot capture inter-atom dephasings.  
Similarly, contrast loss in dynamic environments has been estimated from modelling the imperfect overlap of atomic wavepackets~\cite{Barrett2016b,Roura2014}, but these approaches do not comprehensively consider the microscopic dynamics within the cloud, and therefore do not accurately capture the effects of pulse infidelities, nor how the diversity of atomic trajectories couple with these. 
In turn, Monte Carlo approaches have been developed to describe inhomogeneous atomic clouds and account for pulse infidelities arising from spatially-dependent Rabi frequencies, Doppler detunings and decoherence, while concatenated pulse models have allowed to propagate these effects into interferometer contrast models~\cite{Szigeti2012,Saywell2020,Chiarotti2022}. However, these approaches tend to treat pulses independently and do not capture the microscopic interplay between internal state evolution and
individual non-inertial trajectories.
In this work, we capture the range of these mechanisms by combining a microscopic Monte-Carlo model of the atomic cloud with a semi-classical piecewise numerical evolution of the coupled internal-external dynamics. We note that a similar model structure was proposed in~\cite{Soh2020}, yet did not consider the full range of inertial effects that we do here.\\

\subsection{Modelling concept and implementation}
\label{sec:TwinConcept}

The atom interferometer model that we have developed simulates the coupled dynamics (external motion and internal state dynamics) of each individual atom within the atomic cloud when subjected to an arbitrary sequence of light pulses and arbitrary platform dynamics.
We follow a conventional semi-classical approach that treats external motion classically, while internal dynamics are treated quantum mechanically~\cite{Kasevich1992}.
The state of each atom $i$ is parameterised by six external degrees of freedom, namely its three position $\rbf_i$ and three velocity $\vbf_i$ components, and two internal degrees of freedom, namely the coefficients $c_{G,i}$ and $c_{E,i}$ in the decomposition of its wavefunction $\psi_i$ onto the two relevant atomic states $|G\rangle$ and $|E\rangle$, $\psi_i(t)=c_{G,i}(t)|G\rangle + c_{E,i}(t)|E\rangle$. 
In the context of the $^{87}$Rb two-photon Raman interferometer considered in this work, these states correspond to two hyperfine ground-states, $5^2 S_{1/2}, F=1$ and $5^2 S_{1/2}, F=2$, respectively associated with different external momenta $\mathbf{p}$ and $\mathbf{p}+\hbar\mathbf{k}$ with $\mathbf{p}$ the momentum of the atom.
As the interferometer couples internal and momentum states, these eight degrees of freedom get highly coupled during the interferometric sequence. For instance, the evolution of the internal state under a light pulse depends on the position and velocity of the atom relative to the beam, while the subsequent evolution of the atom's position and velocity is influenced by any momentum kick imparted by the light pulse onto one part of the atomic wavepacket.

The general principle of our simulation is depicted on Fig.~\ref{fig:Fig2}.
At time $t=0$, the cloud is assumed to be in a specified state characterized by a position (e.g. Gaussian), velocity (e.g. thermal Maxwell-Boltzmann) and internal state distribution (e.g. all atoms in the ground state).
All atoms are treated independently in the simulation. 
Performing a Monte-Carlo sampling of the initial cloud, we propagate the external and internal dynamics of each atom along the entire interrogation sequence.
As each atom gets split into wavepackets that experience different non-inertial trajectories, this is achieved, for every atom, by considering all four paths in the interferometer that involve states $|G\rangle$ and $|E\rangle$ (as depicted on the top-right inset of Fig.~\ref{fig:Fig1}), and splitting the sequence, along each of them, in periods of pulse interaction and periods of free propagation.  

\paragraph*{Pulse dynamics -} At each pulse, instantaneous atomic kinematic parameters $\rbf(t)$ and $\vbf(t)$ at the time t the pulse starts, together with the knowledge of external and laser fields, are used to determine the local light fields that the atom experiences (Rabi frequency $\Omega_R(t)$, detuning $\Delta(t)$, phase $\phi_L(t)$), taking into account the laser intensity and phase profiles and the Doppler effect. Those fields are used to numerically propagate the equations of motion describing the interaction between the light and the two-level atom.
In this work, we have considered a $^{87}$Rb two-photon Raman interferometer, for which the equations of motion take the form of the Schrodinger equation,
	\begin{align}
		&i \partial_t \! \left( \begin{matrix} c_G \\ c_E\end{matrix} \right)= \dfrac{1}{2}\left( \begin{matrix} -\Delta(t) & \Omega_R(t)\e^{i\phi_L(t)},       \\ \Omega_R^*(t)\e^{-i\phi_L(t)} & \Delta(t) \end{matrix} \right)\! \left( \begin{matrix} c_G  \\ c_E \end{matrix} \right),
		\label{eq:RWA}
	\end{align}
where the relationships between $\Omega_R(t)$, $\Delta(t)$ and $\phi_L(t)$ and the model input parameters are provided in Appendix~\ref{sec:AppRaman}. 
That way, sensor specifications, such as Raman laser power, Raman beam shapes, sizes or orientations, can be fully accounted for.
As pulses are generally short (a few $\mu s$) compared to the evolution timescale for external degrees of freedom, variations of the atomic position and velocity within each pulse are neglected. Nevertheless, at the end of the pulse, the final quantum state is projected onto the states $|G\rangle$ and $|E\rangle$ and the arm that has received a laser kick ($\pm \kbf$) is allocated a corresponding $\pm\hbar\kbf/m$ velocity increase.

\paragraph*{Free-space dynamics -} Between each pulse, we use the atomic position and velocity right after the preceding pulse as initial conditions to numerically propagate the equations of motion for external degrees of freedom, which read, in the sensor frame,
    \begin{equation}
   \frac{\partial^2 \rbf}{\partial t^2}=\gbf(\rbf)-\abf-2\Ombf\times\frac{\partial \rbf}{\partial t}-\Ombf\times(\Ombf\times\rbf)-\frac{\partial \Ombf}{\partial t}\times\rbf.
    \label{eq:ExtEOM}
    \end{equation}
Here, $\gbf(\rbf)$ denotes the gravity acceleration instantaneously experienced by the atom given its position, while $\abf(t)$ and $\Ombf(t)$ respectively denote the time-dependent acceleration and rotation vector of the interferometer in the Earth's frame. The first term, $\gbf(\rbf)$, is calculated from the knowledge of the instantaneous position of the atom $\rbf$, the local gravity on the ground $\gbf_0$ and the gravity gradient tensor $\mathbf{\Gamma}$ [see details in Sec.~\ref{sec:NavScenarios}]; that way, gravity gradient effects within the interferometer are fully included. The third, fourth and fifth terms in Eq.~\ref{eq:ExtEOM} respectively represent the Coriolis, centripetal, and Euler's acceleration.
As no light coupling is applied between pulses, the time evolution of internal degrees of freedom $c_G, c_E$ amounts to a trivial phase.

This simulation process is handled for all four paths of the interferometer [see Fig.~\ref{fig:Fig2}], allowing to track the individual atomic positions, velocities and internal states along each of them. By accurately replicating the microscopic dynamics taking place in reality, it inherently captures cross-coupling effects as a consequence of their microscopic origins, fully accounting for effects like imperfect overlap of the atomic wavepackets and pulse infidelities.
After the final pulse, atomic populations in each quantum state are inferred and averaged over the initial cloud distribution. As the simulation treats all atoms independently, numerical parallelization is employed. Interferometric fringes can then be obtained by repeating the entire simulation for different values of the chirp rate (see Appendix~\ref{sec:AppRaman}) such that the interferometer phase is scanned.
By considering two vertically separated interferometers, we have straightforwardly extended this model into a gravity gradiometer model. Typical model outputs, as displayed on Fig.~\ref{fig:Fig1}, include interferometric fringes, from which gravity can be extracted, and ellipses, whose fitting allows extraction of gravity gradients. This digital model is capable of providing \textit{ab initio} simulations of gravity gradient measurements from the knowledge of user-specified sensor specifications (including cloud and beams characteristics, sequence details) and conditions of operation (including arbitrary platform dynamics), with no free parameter. 

The model has been quantitatively benchmarked against instrument data collected in laboratory conditions [see for instance Fig. 6 in~\cite{Lellouch2022}]. It was also used to guide the development of the field gravity gradiometer used in Ref.~\cite{Stray2022}. A detailed and thorough comparison with field data would however require exact characterisation, including in-situ analysis, of the instrument and external conditions---which is ongoing work.

\begin{figure*}[t!]
    \centering
    \begin{subfigure}{0.48\textwidth}
        \centering
        \includegraphics[width=\linewidth]{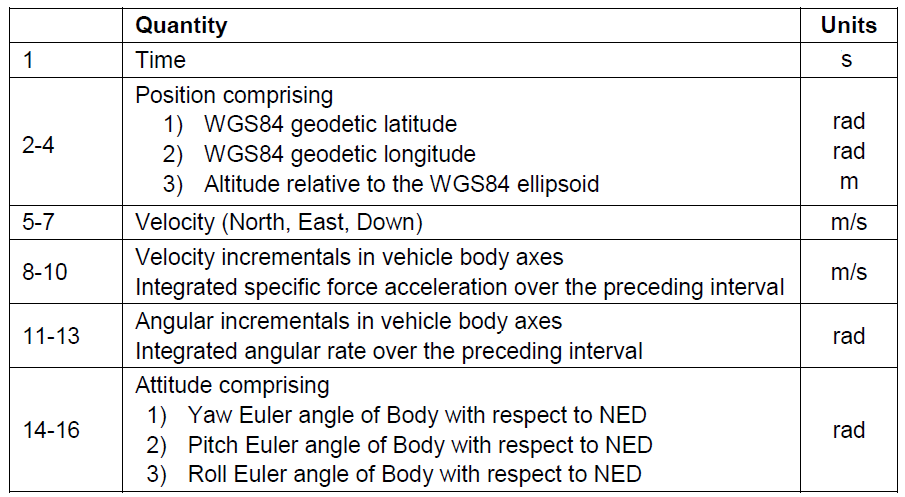}
        \caption{Format of the trajectory data used in this work}
        \label{fig:TrajFormat}
    \end{subfigure}
    \begin{subfigure}{0.48\textwidth}
        \centering
        \includegraphics[width=\linewidth]{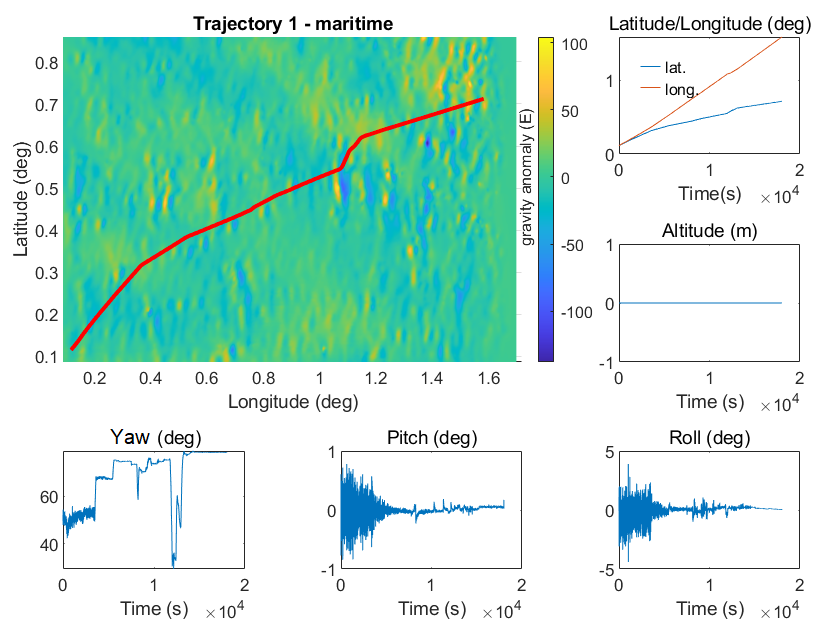}
        \caption{Trajectory 1: ship navigating from sea state 4 to sea state 1.}
        \label{fig:TrajShip}
    \end{subfigure}
    
    \vskip\baselineskip
    \begin{subfigure}{0.48\textwidth}
        \centering
        \includegraphics[width=\linewidth]{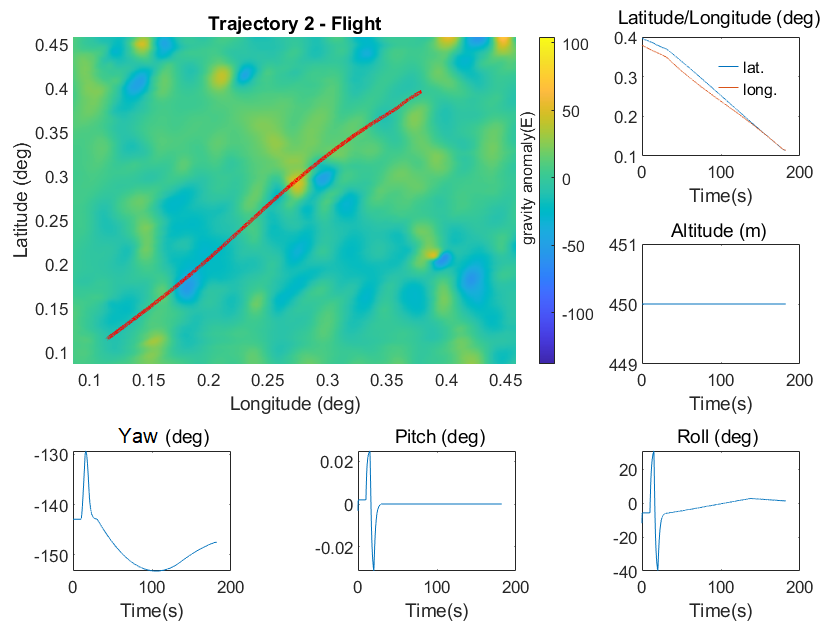}
        \caption{Trajectory 2: constant altitude flight}
        \label{fig:TrajFlight1}
    \end{subfigure}
    \begin{subfigure}{0.48\textwidth}
        \centering
        \includegraphics[width=\linewidth]{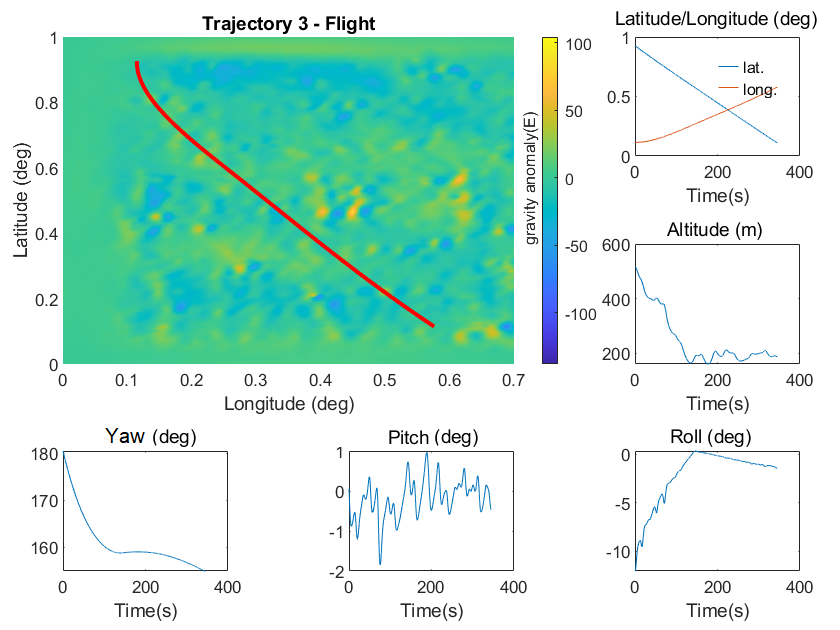}
        \caption{Trajectory 3: terrain-following flight}
        \label{fig:TrajFlight2}
    \end{subfigure}
    
    \caption{Trajectories considered in this work. (a) General format of the trajectory data. This data is sampled at 100Hz. (b)-(d) Depiction of the three trajectories considered (respectively maritime, constant-altitude flight and terrain-following flight). For each trajectory, the main top-left colormap shows the trajectory on the gravity gradient map in a latitude-longitude plane. The five smaller subplots show position (latitude/longitude and altitude) and attitude (yaw, pitch and roll angles) along the trajectories as a function of time.}
    \label{fig:Trajectories}
\end{figure*}

\section{Quantum sensing of gravity gradients along dynamic platform trajectories}
\label{sec:GGMeasurements}

In this section, we consider a quantum gravity gradiometer operating along a variety of platform trajectories representative of different navigation scenarios. By using our model to simulate the sensor measurements, we analyse the impact of platform dynamics and sensor specifications on gradiometric measurements.

\subsection{Navigation scenarios}
\label{sec:NavScenarios}

\paragraph*{Trajectory data - }
We consider trajectories that are representative of both maritime and airborne navigation scenarios. All trajectories are comprised of the data provided in Fig.~\ref{fig:TrajFormat}, sampled at 100Hz.

\begin{figure*}[!t]
\centering
		\includegraphics[width=18cm]{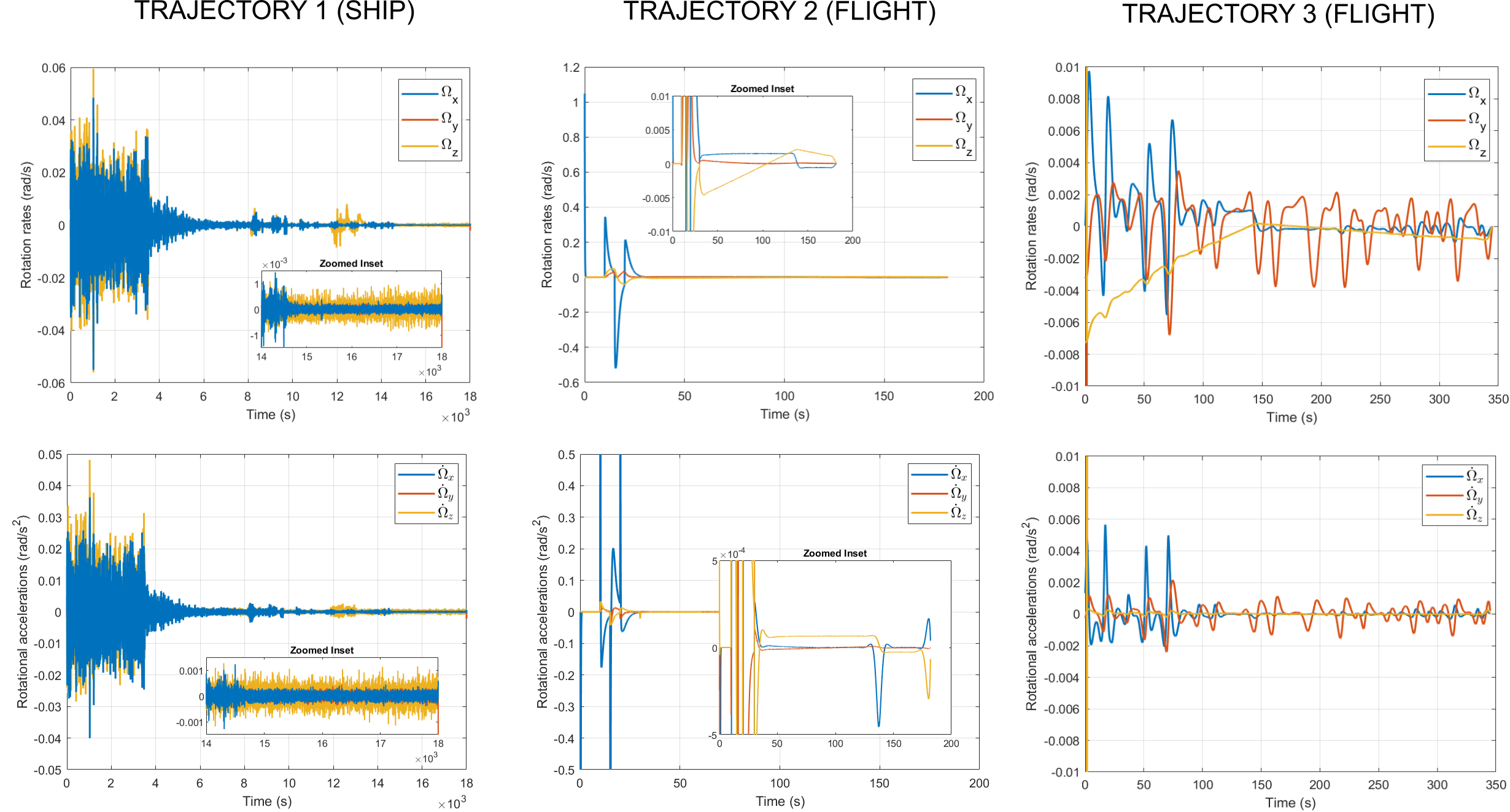}
		\caption{For the three trajectories of Fig.~\ref{fig:Trajectories}, rotation rate (top line) and rotational acceleration (bottom line) in the sensor frame, expressed in a coordinate system attached with the sensor body where x is longitudinal, y is lateral and z is vertical.
      \label{fig:Rotations}}
\end{figure*}

Trajectory 1, depicted on Fig.~\ref{fig:TrajShip}, describes the typical motion of a ship navigating over 5 hours from sea state 4 to sea state 1, as defined in the World Meteorological Organization (WMO) sea state code. Sea state is 4 for $t<4000\mathrm{s}$, 3 for $4000\mathrm{s}<t<6000\mathrm{s}$, 2 for $6000\mathrm{s}<t<15000\mathrm{s}$ (with a few occurrences of a sea state 3 between $8000\mathrm{s}$ and $11000\mathrm{s}$) and 1 for $t>15000\mathrm{s}$. It uses real attitude data (yaw, pitch and roll angles), recorded at 100Hz, to capture the effects of waves, including tilts, rotations and rotation change rates. For simplicity, we have assumed the ship moves at a constant speed of 10m/s and neglected its altitude variations. 
While for a gravity sensor
vertical motion at sea can induce significant phase shifts, it is less relevant to differential gravity gradiometry use cases as it is to high order common to both interferometers, thereby strongly suppressed in
the differential measurement (see Appendix~\ref{sec:AppSubsec2}). 
Centripetal and Coriolis acceleration arising from the ship's frame being non-inertial are fully accounted for in the simulation.

Trajectories 2 and 3, depicted on Figs.~\ref{fig:TrajFlight1} and~\ref{fig:TrajFlight2}, describe two generic examples of airborne platforms, respectively constant altitude and terrain-following.

Figure~\ref{fig:Rotations} further shows, for each trajectory, the rotation rates and rotational accelerations of the platform along the three platform body axes. 

We assume the quantum sensor to be strapdown with the platform, and its position on the platform to be such that the initial position of the cloud in the bottom interferometer coincides with the centre of rotation of the platform. \\

\paragraph*{Gravity environment - }

The gravity environment is encoded in a gravity gradient map, obtained from real data, which has 930 points in latitude and 1576 in longitude, with a grid spacing of $0.0010786$ degrees in both directions. Its values (colorbar  on Fig.~\ref{fig:Trajectories}) correspond to the local $z-z$ component of the gravity gradient tensor $\mathbf{\Gamma}$ after subtraction of its average value on Earth, $3086 \,\mathrm{E}$ ($1E=10^{-9}\mathrm{s}^{-2}$); this so-called gravity gradient anomaly is typically
in the range $[-150E;100E]$. We assume that $\Gamma_{zz}$ does not vary with altitude, and we neglect all other components of the gravity gradient tensor when computing the gravity field at an arbitrary position (latitude, longitude, altitude), namely $\gbf(x,y,z)=\gbf_0+z\Gamma_{zz}(x,y)\mathbf{u_z}$ with $\gbf_0=-9.81\mathrm{ms^{-2}}\mathbf{u_z}$ and $\mathbf{u_z}$ the upward vertical unit vector.

Since the gravity map and trajectories considered in this work are not attached to a specific location on the globe, we have not included the Coriolis force due to the rotation of the Earth in the calculations; its inclusion would amount to introducing an additional term in Eq.~\ref{eq:ExtEOM}.

\subsection{Quantum sensor specifications}
\label{sec:SensorSpecs}

In all the simulations, we have assumed a $1$m-baseline gravity gradiometer, with $10^5$ atoms in each cloud, a total interrogation time $2T=100\mathrm{ms}$ and a $15$ms time-of-flight before the first pulse. 
The situation we consider is that of a well-optimised gravity gradiometer in the laboratory that is put into a navigation environment. In particular, the two counter-propagating lasers are assumed to be generated independently to suppress the effects of parasitic Raman transitions~\cite{Carraz2012}, and their intensity ratio set to cancel out the AC Stark shift. We assume that effective shielding/ magnetic coiling allows to neglect the effects of magnetic fields, and we assume perfectly flat wavefronts. Other sensor parameters including cloud temperature $\theta$ and diameter $d_{\mathrm{cloud}}$, laser power $P$ and beam diameter $d_{\mathrm{beam}}$, have been varied broadly to simulate sensors from lower- to higher- technological readiness. 
In the following, three typical scenarios will be referred to to benchmark the discussion of the results: (i) a "moderate" scenario, representative of current sensor hardware: $\theta=1\mu K$, $d_{\mathrm{cloud}}=5\mathrm{mm}$, $d_{\mathrm{beam}}=15\mathrm{mm}$, $P=0.16W$, (ii) an "enhanced" scenario, with specifications commensurate with current high-performance cold-atom experiments, that could be representative of sensor hardware in the close future: $\theta=1 nK$, $d_{\mathrm{cloud}}=1\mathrm{mm}$, $d_{\mathrm{beam}}=15\mathrm{mm}$, $P=1W$; physically, these parameters correspond to a situation with negligible inter-atom dephasings, where a narrow spread of atomic velocities reduces rotation-induced contrast loss arising from the imperfect overlap between atomic wavepackets, as well-described by the approximate formula $C = \exp(-(k_z \sqrt{2k_B\theta/m})^2 (\Omega_x^2 + \Omega_y^2)T^4)$~\cite{Barrett2016b}; (iii) a "enhanced + mitigated" scenario, where we have assumed that in addition to the sensor specifications of the "enhanced" scenario, additional mitigation techniques, such as a gimbal platform or rotation compensation techniques, would permit to reduce sensor tilts and rotation rates by an arbitrary and tunable numerical factor.

Each gravity gradient measurement is obtained by fitting an ellipse made of 15 data points--each corresponding to one measurement cycle of the atom interferometer~\footnote{About 15-25 shots per ellipse has been shown to be an optimum - large enough to avoid standard errors and small enough to avoid unnecessarily increasing measurement time~\cite{Stray2022}}. These 15 \textit{shots} are successively obtained by running our model with different values of the chirp rate. The laser wavevector is kept unchanged, as we are not looking into effects that can be suppressed by the k-reversal protocol~\cite{Legouet2008}. As the sensor is moving along the trajectory from one shot to the following one~\cite{Phillips2022}, shot repetition rate is an important parameter. To allow investigation of fast interferometric repetition rates, without being limited by the actual duration of the interferometric sequence ($100$ms interrogation time, without adding MOT loading times), we have neglected variations of the sensor kinematic parameters (latitude/longitude, velocity, acceleration, attitude and rotation rates) \textit{within} each shot of the interferometer, allowing us to simulate each ellipse point as if it happened instantaneously. 
This technically allows us to repeat the simulation at interferometric repetition rates $<100$ms even though the interrogation time remains fixed to $2T=100$ms in all our simulations. 
This situation could be representative of using advanced techniques such as interleaving measurements using several sensors or improved detection methods~\cite{Sugarbaker2013}. 
In the following, unless stated otherwise, the \textit{interferometer} repetition rate is 1.5Hz for trajectory 1 (hence a 0.1Hz gravity gradient measurement sample rate) and 15Hz for trajectories 2 and 3 (1Hz gravity gradient measurement sample rate).
On moving platforms, gravity gradient measurements can be impeded by a reduced fringe contrast. In this work, all extracted data points correspond to ellipses with a semi-major axis of above 0.4.

\subsection{Quantum measurements}
\label{sec:GGResults}

\begin{figure}[!t]
\centering
		\includegraphics[width=9cm]{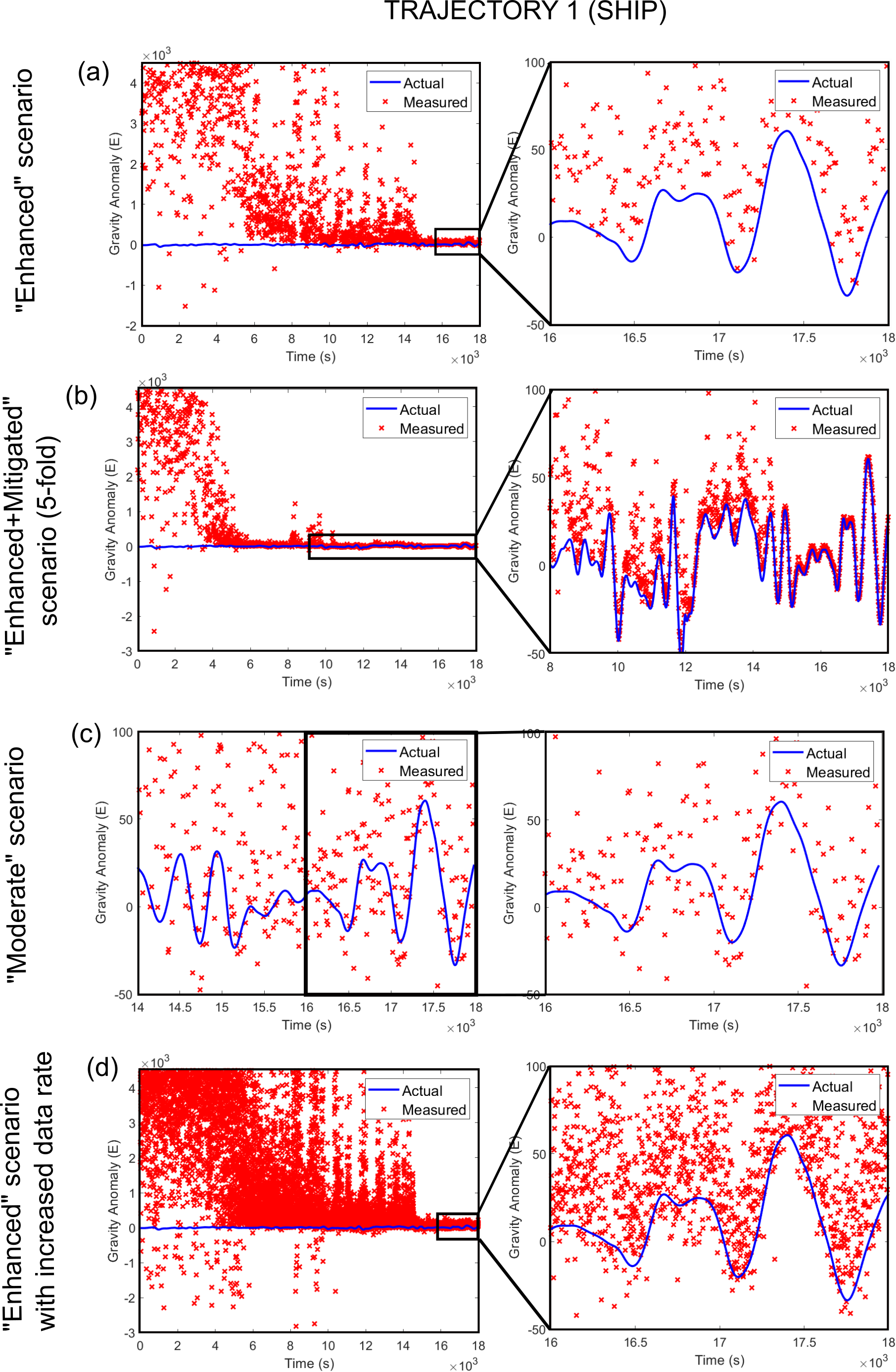}
		\caption{Simulated gravity gradient measurements along trajectory 1, in the: (a) "enhanced" scenario; (b) "enhanced + mitigated" scenario with 5-fold reduction in roll/pitch; (c) "moderate" scenario; and (d) "enhanced" scenario with a higher repetition rate [see main text for details]. The blue line represents the actual gravity anomaly along the trajectory while the red points are the simulated sensor measurements. Selected portions of the trajectory are zoomed in on the right.
      \label{fig:ShipGGmeas}}
\end{figure}
\begin{figure}[!t]
\centering
		\includegraphics[width=9cm]{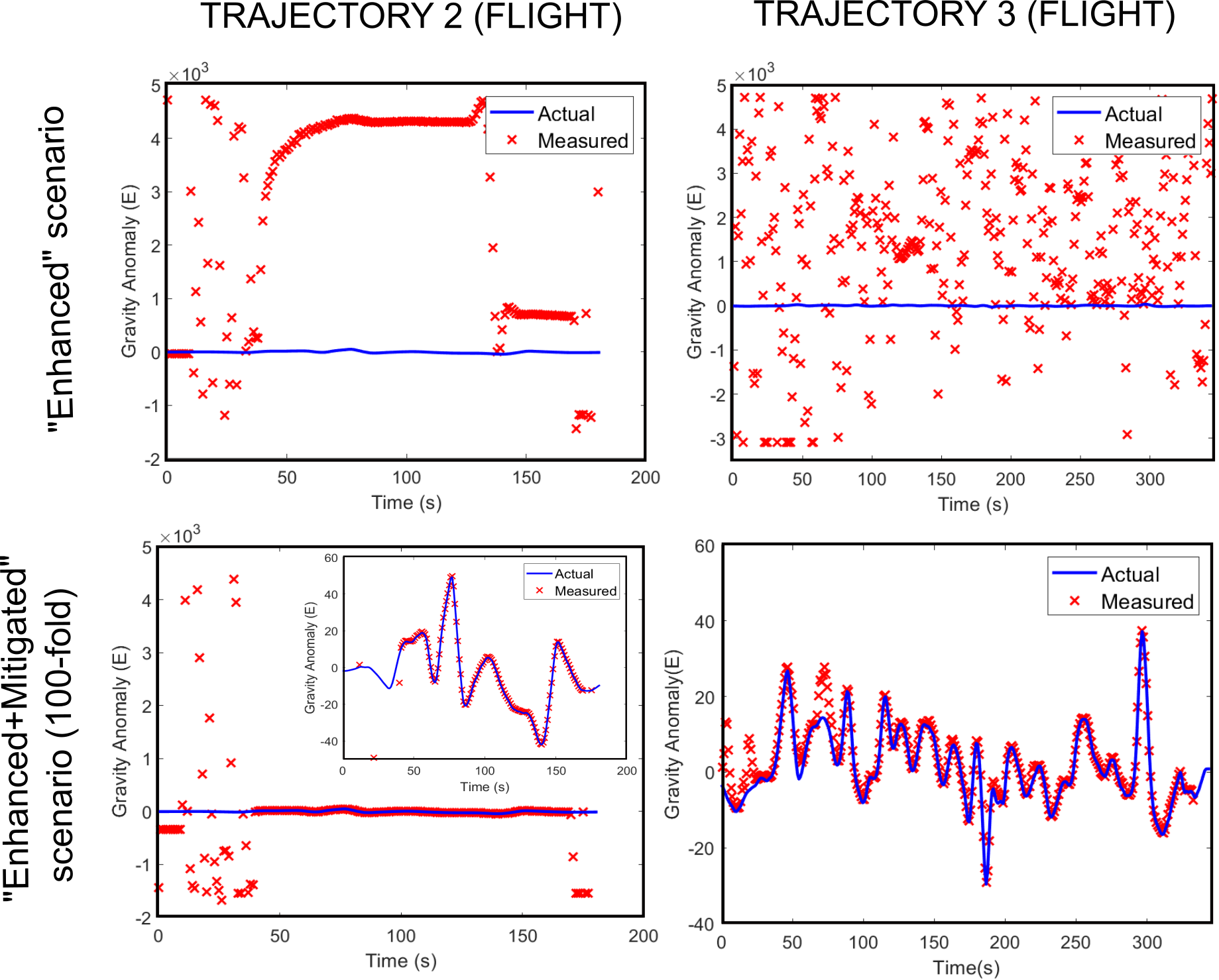}
		\caption{Simulated gravity gradient measurements along trajectories 2 (left column) and 3 (right column) in the "enhanced" scenario (top line) and "enhanced + mitigated" scenario with 100-fold reduction in roll/pitch (bottom line). Data is zoomed in as inset. The blue line represents the actual gravity anomaly along the trajectory while the red points are the simulated sensor measurements.
      \label{fig:FlightGGmeas}}
\end{figure}
\paragraph*{Ship trajectory - } Figure~\ref{fig:ShipGGmeas} displays the simulated gravity gradient measurements along trajectory 1. 
In the "enhanced" specifications scenario (a), the measurements display significant errors all throughout the trajectory, that decrease gradually as the ship goes through calmer sea states. It is is only in sea state 1 ($t\geq 15000$s), that the gravity gradient signal is not completely written out by noise [see inset]. Conducting additional simulations while isolating out different contributions of the ship motion has shown that these errors were imputable to both instantaneous rotation rates and rotational accelerations. 
Both significantly contribute to the interferometer phase shift, leading to a systematic error in the estimated gravity gradient that, due to the rapid change in these quantities between consecutive gravity gradient measurements, appears as noise.
Importantly, we note that these manifest themselves as non-Gaussian noise, and induce in particular a positive bias on gravity measurements, as can be understood from a simple approximate calculation of the rotation-induced phase shift [see Appendix~\ref{sec:App1}]. 
In turn, static tilts and ship transverse acceleration were found to have a limited effect, although in sea state 4 we do start to see effects of pulse infidelities arising from the atoms shifting off-centered in the beam as a result of tilt and centripetal force. Those effects would be more pronounced if assuming the sensor was positioned further away from the platform centre of rotation [see discussion in Appendix~\ref{sec:AppSubsec2}].
Finally, in rough sea states, high noise in the elliptical signal and large phase shifts approaching 0 or $\pi$ give rise to systematic ellipse fitting errors that are a contributing factor to the observed non-Gaussian noise. It is possible to circumvent this via active control of the ellipse phase, which is routinely performed in static differential interferometers but not yet demonstrated in the public domain at sea, so in order not to underestimate the implications of these effects, we have not employed such techniques here.
Overall, the observed noise levels range from several $10^3 E$ in sea state 4 to several $10E$ in sea state 1.  
To mitigate these, we have simulated an "enhanced + mitigated" scenario where roll and pitch angles have been reduced by a factor of 5. This scenario (b) shows a significant performance improvement, evidencing measurements capable of resolving the gravity gradient variations of the map throughout sea states 1 and, to a lesser extent, 2, and a 1$\sigma$ noise level of the order of $25E$ in sea state 2 and $\approx 1E$ in sea state 1.
We have also investigated the effect of reducing sensor technological readiness through considering the "moderate" specification scenario. As shown in Fig.~\ref{fig:ShipGGmeas} (c) in the case of calm sea states, it evidences an increase of noise compared to the "enhanced" scenario as a result of cross-coupling effects arising from atomic cloud inhomogeneities (hotter, larger cloud). Interestingly though, this additional noise arising from inter-atomic cross-couplings fundamentally differs from the previously described inertial errors in the sense that it manifests itself as Gaussian noise. In rougher sea states 3 and 4, we have found that with ``moderate'' sensor specifications, the contrast reduces to a level which limits the accuracy of current ellipse extraction techniques.
Finally, Fig.~\ref{fig:ShipGGmeas} (d) shows simulations of the "enhanced" scenario with a higher interferometer repetition rate of 10Hz per shot (hence a 0.67Hz gradiometry sample rate). We observe that measurements are slightly noisier, which here does not arise via a reduced interrogation time, but rather via the fact that high data-rate measurements are more sensitive to high-frequency rotational noise from waves. Indeed, when using a lower data-rate, changes in inertial quantities between consecutive ellipse points result in these points sitting on different ellipses - producing a noisier ellipse, yet whose fitting smooths out these effects.\\

\paragraph*{Flight trajectories - }Figure~\ref{fig:FlightGGmeas} displays the simulated gravity gradient measurements along trajectories 2 and 3. Even in the "enhanced" scenario (top plots), the gravity gradient measurements are severely deteriorated. For trajectory 2, we find that the significant systematic shift occurring for $t>50$s is due to variations of the roll angle, which create a rotational phase shift $\propto \Omega_x^2$ [see Appendix~\ref{sec:App1}] in the interferometer that largely exceeds the gravity-gradient induced phase shift; we find that the measurement curve on Fig.~\ref{fig:FlightGGmeas} (top left) closely follows the variations of the roll angle, as observed through $\Omega_x$ on Fig.~\ref{fig:Rotations}, so that the sensor actually measures a rotation rate instead of a gravity gradient. For $t<50$s, sensor tilts, accelerations and rotations are all responsible for the observed noise, each resulting in significant errors even when taken out individually. Similar observations are made for trajectory 3, where large systematic shifts are imputed to variations of the pitch angle primarily and, for $t<100$s, to large sensor tilts in the roll direction. For both trajectories, the "enhanced + mitigated" scenario, which we have simulated here with a 100-fold reduction in roll and pitch angles, shows a highly improved performance, with a 1$\sigma$ gradiometry residual noise below $1E$ over most of the trajectories. We note that due to the level of errors observed in the ``enhanced'' scenario, the ``moderate'' scenario has not been simulated in flight trajectories.\\

\section{Integration with a map-matching navigation filter}
\label{sec:MM}
\begin{figure*}[!t]
\centering
		\includegraphics[width=18cm]{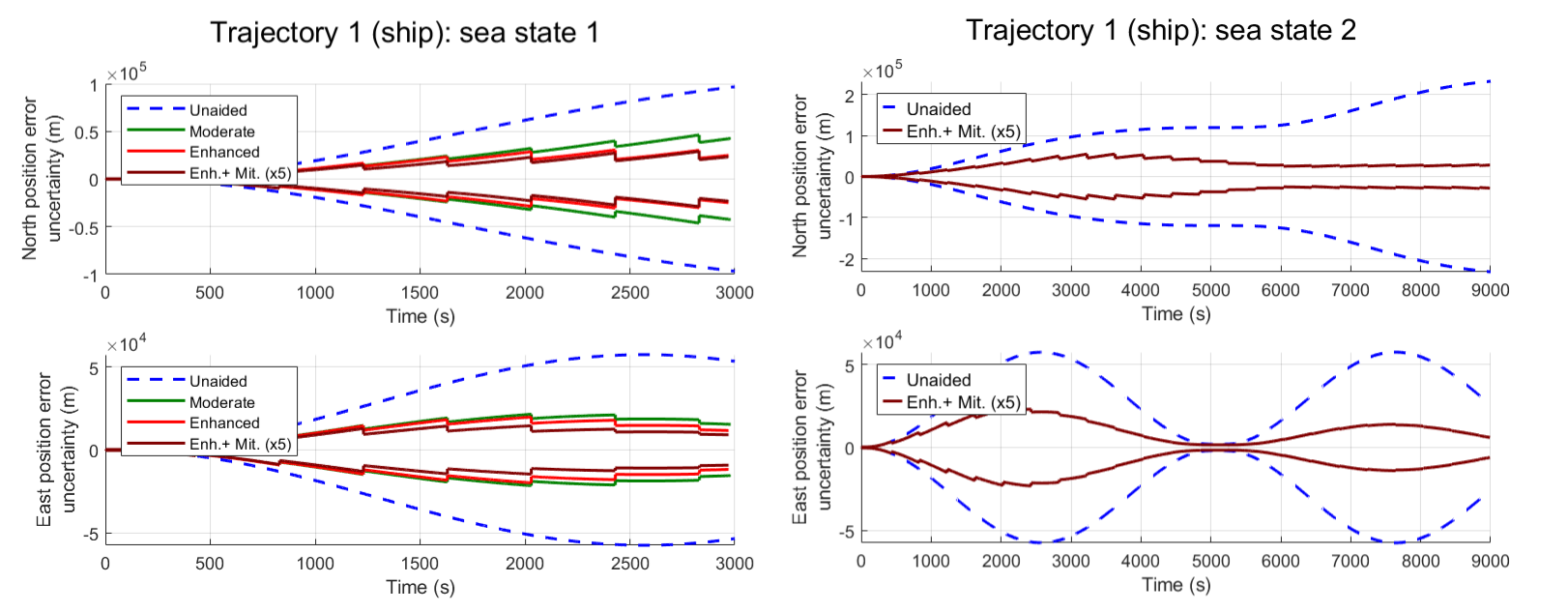}
		\caption{North (top line) and East (bottom line) position error uncertainty for portions of trajectory 1 respectively corresponding to: (left column) sea state 1 (portion $15000\mathrm{s}<t<18000\mathrm{s}$ of trajectory 1); (right column) sea state 2 (portion $6000\mathrm{s}<t<15000\mathrm{s}$ of trajectory 1). On each subplot, the dashed blue line represents the unaided case (INS only) while the various solid coloured lines correspond to different levels of sensor specifications. The "enhanced + mitigated" scenario is implemented here with 5-fold reduction in tilts/rotations.
        \label{fig:FigShipMM}}
\end{figure*}
In this section, we demonstrate the integration of gradiometry digital measurements within an error-state Integrated Navigation filter. 
For the sake of simplicity here, we will only be interested in 2D position fixing.
Gradiometry measurements are first processed through a map-matching algorithm to generate periodic horizontal position estimates. This estimate is then passed to the navigation filter, which provides position, velocity and attitude corrections to the inertial navigation solution derived from the Inertial Measurement Unit (IMU). 
In this work, we are considering tactical/platform grade IMUs, which are among the most commonly used grades for high-capability navigation, such as for GPS holdover, across a broad range of navigation use cases --although not state-of-the-art (e.g. strategic or navigational grade). Therefore, we have assumed the following IMU errors: for trajectory 1 (ship - platform grade), an accelerometer bias of $25\mu$g, scale factor error of 25ppm and misalignment of $11.6\mu$rad, and a gyroscope bias of $0.0023^o$/hr, scale factor error of 5ppm and misalignment of $9.3\mu$rad; for trajectories 2 and 3 (flight - tactical grade), an accelerometer bias of 3mg, scale factor error of 1000ppm and misalignment of 0.5mrad, and a gyroscope bias of $10^o$/hr, scale factor error of 1000ppm and misalignment of 0.5mrad.
All simulations start with 30s of Transfer Alignment (TFA) which prime the navigation solution and during which gradiometric measurements are not utilized to initialise the system position, velocity and attitude states. A TFA initialisation error of 100m North and 100m East is assumed in all cases.

The map-matching filter considered here is a particle filter.
An overview of the algorithm, its implementation details and key tuning parameters is provided in Appendix~\ref{sec:AppPF}.
For each trajectory, we have first used synthetic data with various noise levels to verify that the algorithm is broadly functioning correctly and resulting in stable performance. This allowed us to derive suitable tuning of key parameters of the filter, including the number of measurements accumulated before resampling and the sensitivity of the particle weighting to the innovation.
Optimal tuning of the filter hyperparameters is a subtle problem that is \textit{a priori} highly specific to each trajectory (platform dynamics, terrain characteristics) and sensor capability (measurement noise, data rate). 
As this work focuses on demonstrating successful integration of quantum and map-matching capabilities, and assessing key filter input requirements, rather than maximizing ultimate navigation performance, we have prioritized generic tunings that ensure covariance consistency across all measurement capabilities over platform- and sensor-specific optimization.
The only tuning parameter that we have allowed to vary when varying the sensor capability and the trajectory type is the particle filter probability density function (PDF) used for weighting particles based on their innovation [see Appendix~\ref{sec:AppPF}]. 
In brief, the narrower the particle filter PDF, the more responsive the navigation filter will be to measurement updates, yet the filter will also be more at risk to become covariance inconsistent whenever the observed position measurement error lies outside the expected uncertainty bounds. 
To mitigate this, particle weighting must be based on measurement uncertainty.
From preliminary simulations using synthetic data, we have found that tuning the particle filter PDF standard deviation to 2$\sigma$, with $\sigma$ the observed gradiometric measurement noise, allowed to guarantee filter stability across all capabilities considered here. 
As a result, when performing map-matching simulations on trajectory 1 where the measurement noise is strongly varying across sea states, we have had to split the trajectory into portions corresponding to different sea states and tune the filter to the observed measurement noise in each of them, as would be the case if the tuning was dynamic based on knowledge of the sea state.\\
\begin{figure*}[!t]
\centering
		\includegraphics[width=18cm]{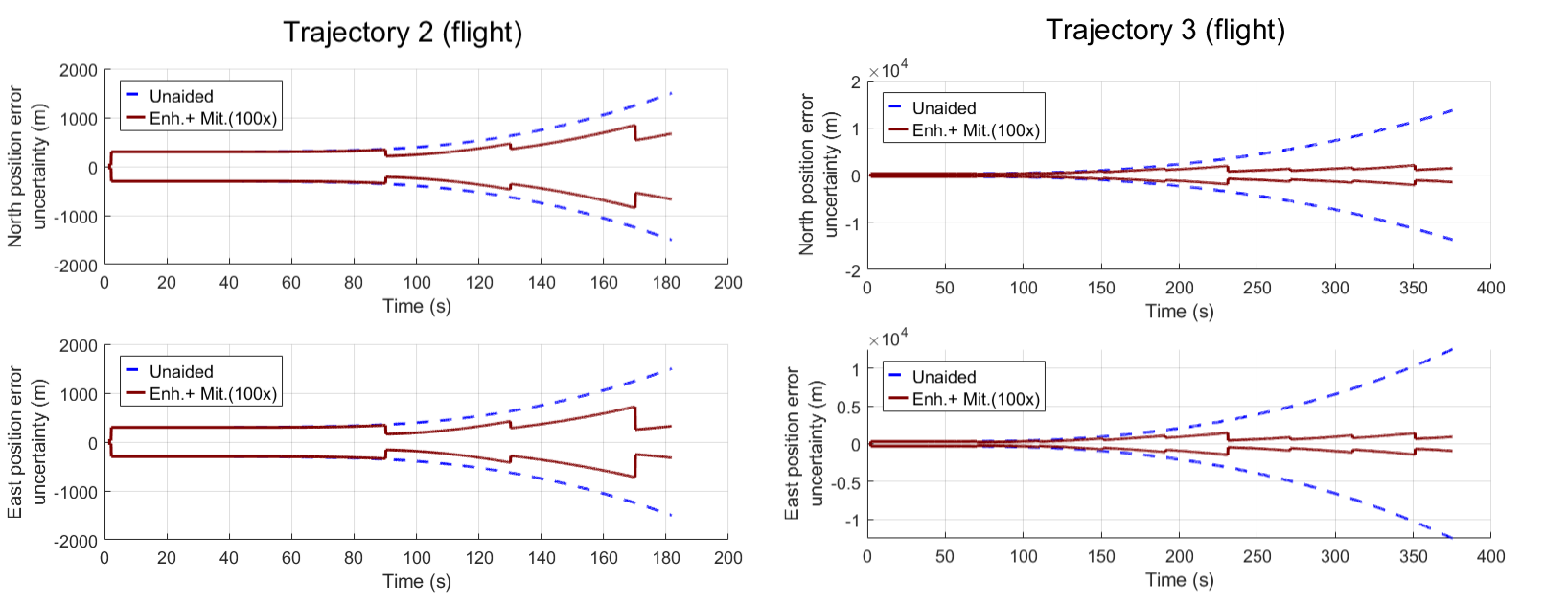}
		\caption{North (top line) and East (bottom line) position error uncertainty for flight trajectories 2 (left column) and 3 (right column), in the non-aided case (dashed blue line) and "enhanced + mitigated" scenario implemented here with 100-fold reduction in tilts/rotations (solid brown line).
        \label{fig:MapMatchFlight}}
\end{figure*}
Figure~\ref{fig:FigShipMM} displays map-matching results for sea states 1 and 2 using the simulated gravity gradient measurements of Sec.~\ref{sec:GGResults}. It shows the North and East position error uncertainty as a function of time for various sensor capabilities.
While in the non-aided (INS only) case, position errors grow as a function of time, periodic position updates (here, every 400s) when using map-matching allow to combat covariance growth. We note that the periodicity observed in the East position error is a well-known phenomenon known as Schuler oscillations - a characteristic 84-minute cyclic response of INS to incorrect initialization and inertial sensor errors.
In sea state 1 [Fig.~\ref{fig:FigShipMM}, left], we find that all sensor capabilities considered show some improvement over the non-aided case, reducing the growth of position errors by at least a factor of 2, while achieving bounding (in the sense that periodic position corrections at least compensate for the error growth over the preceding resampling interval) in the "enhanced" and "enhanced + mitigated" capabilities.
The limited benefits observed while transitioning from the "enhanced" to the "enhanced + mitigated" capability can be attributed to sub-optimal filter tuning, as the particle filter has been configured to privilege robustness, but at a cost of sensitivity to measurements. On the one hand, the $2\sigma$ criterion used when weighting particles effectively prevents inadvertent deletion of particles as would arise if weighting them close to zero in an extreme measurement event, but is excessively cautious in the "enhanced + mitigated" case where the measurement noise is strongly suppressed. On the other hand, the long resampling interval (400s) generally ensures that enough measurements are processed to achieve convergence in all scenarios, but could be shortened in those where gradiometric measurement uncertainty is low to allow more frequent position fixes, helping to mitigate covariance growth. 
A careful optimization of these parameters would however need to take into account not only the underlying gradiometric measurement uncertainty, but also the distance of terrain traveled.
Therefore, while these results demonstrate the merit of gravity gradient aiding in sea state 1 environments, even with current quantum sensing capabilities, they also underline the importance of having an end-to-end simulation tool that could underpin future, platform- and sensor-specific optimization of the navigation solution as a whole.

In sea state 2, the significant errors in the "moderated" and "enhanced" measurements did not permit to derive suitable filter tuning. Errors significantly larger than the underlying noise of the sensor -essentially causing the error to be highly non-Gaussian- pose a significant challenge for map-matching algorithms, which assume a Gaussian error. To prevent these errors from causing filter instability, we have had to detune the filter, resulting in uncertainties that were too large to meaningfully improve the positioning accuracy. 
In turn, a noticeable step in performance was observed when transitioning to the "enhanced + mitigated" scenario, implemented here with 5-fold reduction in tilts/rotations. The latter [see Fig.~\ref{fig:FigShipMM}, right] displays significant reduction and successful bounding of position errors over time, evidencing the crucial need for successfully mitigating platform dynamic non-Gaussian errors. 

Similar conclusions were obtained for sea states 3 and 4, as well as for flight trajectories 2 and 3, where, due to large systematic errors, only the "enhanced + mitigated" scenario could be used to exercise the map-matching algorithm. However, due to the large errors present, we have found that more severe mitigation was required. 
Fig.~\ref{fig:MapMatchFlight} shows map-matching results for trajectories 2 and 3, assuming a 100-fold reduction in tilts/rotations. With position errors remaining bounded over time, it shows that such "enhanced + mitigated" gradiometric measurements at 1Hz are sufficient to allow efficient position fixing even in high-dynamic platforms.

\section{Discussion}
\label{sec:discussion}

Our results have evidenced that quantum gradiometry can successfully be used for navigation aiding in both air and ship use cases, even in the presence of important noise, and can turn an unbounded trajectory into a stable trajectory when aiding an INS. We have found that capability is however critically dependent on the successful mitigation of centripetal and/or Coriolis effects to avoid significant non-Gaussian errors in the gradiometry measurement, that cause issues upon integration with map-matching filters. This is particularly prevalent in air trajectories that have high dynamics but also critical for maritime navigation in rough sea states. 

Our results have shown that mitigation strategies, whereby rotation rates are reduced by an arbitrary factor to mimic an externally or internally stabilised sensor, such as via a gimble or active beam control respectively, are effective, resulting in stable trajectories. 
It is possible to derive simple, rough criteria on residual sensor tilt $\alpha$, rotation $\Ombf$ and rotation rate $\dot{\Ombf}$ to be met for effective gimballing. To do so [see Appendix~\ref{sec:App1}], we impose that the interferometric phase shift induced by these effects remain smaller than the gravity-induced phase shift associated with a small feature on the map, which gives:  
\begin{eqnarray}
    \alpha & < &  3.3^o \nonumber\\
    \Omega_{S_{x,y}} & < & 7.10^{-5}s^{-1} \nonumber\\
     \dot{\Omega}_{{S}_{x,y}} & < & 7.10^{-4}s^{-2}  
     \label{eq:Criteria}
\end{eqnarray}
These criteria are found to be generally consistent with what is observed in Sec.~\ref{sec:GGResults} and with the levels of successful mitigation evidenced for the various platforms considered in this work given the rotation rates and rotational accelerations shown on Fig.~\ref{fig:Rotations}.

However, especially on future air and compact marine platforms where the use of a gimbal sensor may not be practical, a strap-down sensor might be required. Different areas to progress on the mitigation of inertial errors in strap-down sensors may be pursued in the future, including: (i) Developing resilient hardware, such as driving the interferometer using composite~\cite{Saywell2018,Saywell2020,Saywell2023} or polychromatic~\cite{Lellouch2022} pulses, which have shown promises to enhance the sensor’s resilience to external field and atomic cloud inhomogeneities. Such pulses could be tailored to mitigate the effects induced by tilts, accelerations and rotations using robust and optimal quantum control techniques~\cite{Glaser2015}. High-bandwidth sensors~\cite{Rakholia2014,Lee2022,Adams2023} operating at small interrogation times could also help reduce the impact of centripetal and Coriolis effects.
(ii) Post-processing techniques to compensate systematic errors out of the gradiometry measurements using additional information such as IMU data. While hybridization between a classical IMU and an atom interferometer has been demonstrated in other navigation contexts~\cite{Tennstedt2023}, a concept for gravity retrieval in an atom interferometer hybridized with an electrostatic accelerometer has been proposed in~\cite{Zahzam2022} and separation between rotation and acceleration signals in an atom interferometer hybridized with a classical sensor has been demonstrated in~\cite{Darmagnac2024}. To be effective in regimes where cross-couplings are significant, such compensation methods would require as accurate an interferometer model as the one presented here. They would also need overcoming potential issues related to non-linearity in fringe reconstruction or underpinning sensor hardware.

While we have focused on errors induced by dynamic conditions of operations, other external factors, such as magnetic fields, may induce similar non-Gaussian errors. Effective suppression of DC magnetic effects has been demonstrated in field gravity gradiometers via the use of passive and active shielding, magnetic coiling and the implementation of techniques such as interleaving measurements~\cite{Stray2021}. Their implementation on moving platforms, especially flight applications, is mostly dependent upon meeting SWAP constraints. Interestingly, as map-matching algorithms are highly tolerant of Gaussian noise, reducing quantum sensor SWAP could still bring benefits for navigation despite coming with increased noise from inter-atom cross-couplings.

More generally, couplings between the dynamic environment and common systematic effects in atom interferometers, such as light-shifts, Zeeman shifts or wavefront aberration, may induce additional errors that would need further scrutiny. As we have considered a well-optimised gravity gradiometer, with these effects efficiently mitigated in the lab, which is then put into a navigation environment, we would expect these systematic effects to be broadly canceled out. However, possible residual contributions arising from the interplay with the dynamic environment (e.g. dynamical light-shifts, non-inertial atomic motion in aberrated wavefronts...) could give rise to non-trivial errors. While these would not affect the general conclusions of this work, such as the distinction between Gaussian and non-Gaussian errors, they may impact the exact performance, and a detailed analysis of these couplings may constitute the object of a follow-up study. 

Our integration of a quantum sensor model with a navigation filter has also demonstrated how gradiometry measurement errors impact navigation filter requirements. Rather than seeking to suppress/compensate these errors, an alternative approach would be to make map-matching algorithms more tolerant to them, such as by removing the underlying Gaussian noise assumption. More generally, while filter tuning and quantum sensing error mitigation are traditionally considered separately, our results have highlighted how the optimization of the navigation and the quantum sensor should be considered together. To make the filter as responsive as possible to measurements without risking to become covariant inconsistent, optimal filter tuning should take into account the uncertainty in quantum measurements, which strongly depends on sensor specifications and platform dynamics. Meanwhile, finding the optimal resampling time that most efficiently combats covariance growth is expected to be heavily influenced by the measurement uncertainty, the map roughness and the platform velocity (which influence the chances for a unique position fix), but also by the gradiometer measurement rate; optimally tuning the latter involves a non-trivial trade-off between ensuring enough measurements have been processed to achieve convergence, and noisier quantum measurements/risk of duplicated measurements.
In that respect, integrating a quantum sensor model with a navigation filter represents a first critical step on the way to finding the optimum measurement configuration and filter tuning that would be unique to the platform kinematics, map roughness, quantum sensor specifications, and depend on end-user performance requirements.\\

\section{Conclusion}
\label{sec:CCL}

In this work, we have demonstrated the successful integration of a high-fidelity microscopic model of a cold-atom gravity gradiometer with a navigation filter into an end-to-end simulation tool of a quantum-enhanced navigation solution as a whole. 
Through this integration, we have identified key challenges that must be addressed to advance the use of quantum sensors in navigation, particularly the emergence of non-Gaussian errors in quantum sensor measurements due to the platform dynamics.
We have derived mitigation requirements for these errors and shown that, under these conditions, aiding navigation via map matching using gravity gradiometry results in stable trajectories. 
This work, and the developed model, will allow development of filters for advancing early map-matching capabilities on maritime platforms.
Furthermore, it constitutes a first significant iteration towards a full digital twinning capability that would support putting real sensors into trials and enabling joint, platform-specific optimization of both the quantum sensor and navigation filter.
Beyond navigation applications, our atom interferometer model has the potential to address current modelling gaps in research endeavours ranging from practical quantum sensing to fundamental physics applications~\cite{badurina2019aion,canuel2019}. With quantum sensors moving into increasingly complex environments and within increasingly complex systems, comprehensive modelling approaches establishing a quantifiable link between sensor specifications, conditions of operation and application performance could bring significant value in best targeting the needs of end users of the technologies.\\

\noindent \textbf{Acknowledgements:}
The authors would like to thank MBDA for collaborating on model integration and supporting the map-matching simulations, and Adam Seedat for insightful discussions on maritime trajectories. The authors acknowledge funding from the Engineering and Physical Sciences Research Council (EP/T001046/1) and Ministry of Defence, as part of the UK National Quantum Technologies Programme.\\

\noindent \textbf{Author contributions:} SL created the modelling framework, performed the simulations and wrote the article. SL and MH both conceived and worked together on the study and the development of the manuscript.\\

\noindent \textbf{Data availability statement:} All data and code supporting the findings of this study can be available from the corresponding author upon reasonable request.\\

 \noindent \textbf{Competing Interests:} The authors declare that they have no competing interests. \\
 
\bibliography{biblioSLBirmi}

\appendix

\section{Two-photon Raman transitions and application to the $^{87}$Rb atom interferometer considered in this work}
\label{sec:AppRaman}

\begin{figure}[!h]
\centering
		\includegraphics[width=8cm]{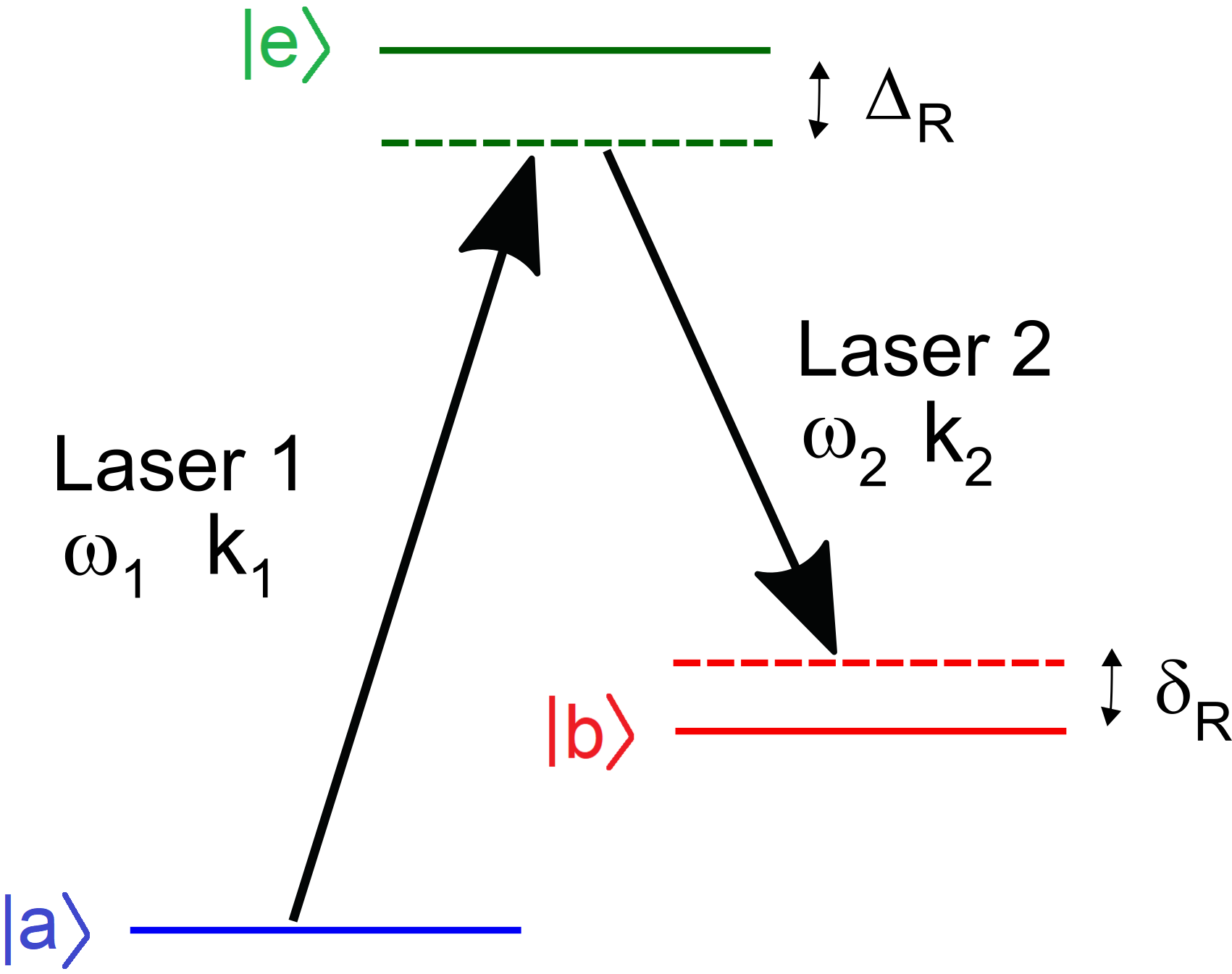}
		\caption{Schematic depiction of a stimulated Raman transition.
      \label{fig:Raman}}
\end{figure}
In a stimulated Raman transition, two hyperfine ground-states $|a\rangle$ and $|b\rangle$ are coupled by two laser beams whose frequencies $\omega_1, \omega_2$ are chosen to be close to the resonant frequencies between those states and a common excited state $|e\rangle$, $\omega_1\approx \omega_{ae}, \, \omega_2\approx \omega_{be}$; the detunings $\Delta_R$ and $\delta_R$ [see Fig.~\ref{fig:Raman}] are referred to as the single- and two-photon detuning.
Defining $\Omega_j=-D\sqrt{2I_j/c\epsilon_0}/2\hbar$ the Rabi frequency associated with laser j, with $I_j$ the intensity of laser j, $c$ the speed of light, $\epsilon_0$ the vacuum dielectric permittivity and $D$ the transition dipole matrix element, one can show that if $\Delta_R \gg \Omega_1, \Omega_2$, state $|e\rangle$ remains mostly unpopulated and the dynamics of the system reduces to that of a two-level system [$|a\rangle$, $|b\rangle$] governed by the 2x2 matrix
	\begin{align}
		\dfrac{1}{2}\left( \begin{matrix} -\Delta & \Omega_R\e^{i\phi_L},       \\ \Omega_R^*\e^{-i\phi_L} & \Delta \end{matrix} \right)\! 
		\label{eq:RWAapp}
	\end{align}
with:
\begin{itemize}
    \item $\Omega_R = \sqrt{\Omega_\textbf{eff}^2 + \Delta^2}$ where the detuning $\Delta$ is defined below and the effective Rabi frequency $\Omega_\textbf{eff}$ is given by
    $\Omega_\textbf{eff} = 2\frac{\Omega_1\Omega_2^*}{\Delta_R}$~\cite{Berman1997}.  
    In this work, the single-photon detuning $\Delta_R$ is set to 2.1GHz, and the Rabi frequencies $\Omega_{1,2}$ depend on the instantaneous atomic position through the relation $\Omega_j=-D\sqrt{2I_j(\rbf)/c\epsilon_0}/2\hbar$ with $I_j(\rbf)$ the local intensity of laser j at the instantaneous atomic position $\rbf$.  
    \item $\Delta=\delta_{AC}-\delta_R-\mathbf{k}.\vbf$ where $\delta_{AC}=\frac{|\Omega_1|^2}{\Delta_R}-\frac{|\Omega_2|^2}{\Delta_R}$ is the differential AC Stark-shift and $-\mathbf{k}.\vbf$ is the Doppler frequency shift due to the atom's motion, with $\mathbf{k}=\mathbf{k_1}-\mathbf{k_2}$ the effective wavevector and $\vbf$ the instantaneous atomic velocity. 
    \item $\phi_L$ is the differential phase between the two Raman beams.
\end{itemize}

In this work, we assume the two laser beams to come from separate, counter-propagating monochromatic lasers, whose frequencies are chosen so that an atom at rest is on resonance, $\delta_R=0$. 
A linearly increasing chirp $\alpha_0 t$ is further applied to $\omega_1$ as the cloud free-falls to compensate for the gravity-induced Doppler shift, while varying the chirp rate $\alpha_0$ is used to scan the interferometric fringe.
We assume both lasers to have the same Gaussian intensity distribution, and the intensity ratio between the two beams to be set so that $\delta_{AC}=0$. For simplicity, we assume perferctly flat laser equiphases, $\phi_L=\mathbf{k}.\rbf$.\\

In practice, a real $^{87}$Rb atom features a more complex atomic state manifold. Due to the hyperfine splitting of the $5^2 P_{3/2}$ excited state, the stimulated Raman transition is no longer strictly a three-level process, as magnetic sub-levels create multiple possible routes for the stimulated Raman transition to happen via any intermediate $m_F$ state. In that context, the effective Rabi frequency $\Omega_\textbf{eff}$ of the stimulated Raman transition between the hyperfine ground-states $F = 1$ and $F = 2$ is obtained by summing the contributions of all possible transitions involving different intermediate $m_F$ states. Whichever transitions are allowed depends on the polarization of the Raman lasers. Assuming both beams have the same polarisation, this leads to $\Omega_\textbf{eff} = 2\frac{\Omega_1\Omega_2^*}{\Delta_R}\frac{\sqrt{4 - m_F^2}}{12}$ with $m_F$ the magnetic quantum number in the ground-state~\cite{Weiss1994}. This, with $m_F=0$ (most interferometers operate on magnetic-insensitive states), is the expression we have used in this work. 

\section{Estimation of systematic effects and mitigation requirements}
\label{sec:App1}

In this appendix, we compute simple estimates for the systematic errors arising from operating in a dynamic environment, and use them to derive requirements to be met by a gimballed sensor to allow successful mitigation of these systematic effects.

We consider here a simplified single-atom picture of the interferometer (which approximates well the ‘enhanced’ scenario considered in this work where atomic cloud inhomogeneities remain small). The motion of the atom is described by Eq.~\ref{eq:ExtEOM}~\footnote{Consistently with the rest of this paper, we still neglect here the Coriolis force due to the Earth's rotation.}. As the associated Lagrangian is quadratic in position and velocity, the interferometric phase shift can be computed using the mid-point-theorem~\cite{Antoine2003,Overstreet2021}, $\Phi=\Sigma_{i=1}^3 (\kbf^{(I),i}-\kbf^{(II),i}).\bar{\rbf_i}$ where $\kbf^{(I),i}$ (resp. $\kbf^{(II),i}$) is the effective laser wavevector at pulse $i$ along arm I (resp. II) of the interferometer and $\bar{\rbf_i}=(\rbf^{(I)}_i+\rbf^{(II)}_i)/2$ the average atomic position at pulse $i$.
The latter can be obtained by solving Eq.~\ref{eq:ExtEOM}, where $\abf$ and $\Ombf$ represent the acceleration and rotation vector of the sensor in the Earth's frame. In the following, we assume that the sensor rotation vector varies at most linearly
with time, $\Ombf(t)=\Ombf+\dot{\Ombf} t$ and we neglect variations of the fields $\gbf$ and $\abf$ experienced by the atom during the interferometric sequence. This notably means that for simplicity in this Appendix, gravity gradient effects \textit{within each individual interferometer}, which are captured by our numerical model via the term $\mathbf{g(\mathbf{r})}$ in Eq.~\ref{eq:ExtEOM}, are neglected in this calculation; a discussion of their effects is however provided in Sec.~\ref{sec:AppSubsec2}. Following the approach proposed in~\cite{Darmagnac2024}, that we extend here to arbitrary rotations, we seek an approximate polynomial solution to Eq.~\ref{eq:ExtEOM} for $\rbf(t)$, truncated to third order in time. We denote respectively $\rbf_1=(x,y,z)$, $\vbf_1=(v_x,v_y,v_z)$ and $\mathbf{a}=\gbf-\abf$ the atom's initial position, velocity and acceleration at the time of the first pulse; all vectors are defined in the sensor frame and expressed in a coordinate system attached with the sensor body where x is longitudinal, y is lateral and z is vertical. 
We assume the lasers to be strapdown with the sensor so that their wavevectors $\kbf_{i}$ rotate along with it and the effective wavevector $\kbf=\kbf_1-\kbf_2$ remains fixed in the sensor frame. Denoting $\kbf=k_z\mathbf{e_z}$ with $\mathbf{e_z}$ the vertical unit vector in the sensor's frame, and rewriting $\mathbf{\Omega}\equiv\Ombf$ for convenience, the phase shift is obtained as: 
\begin{eqnarray}
     \Phi & = & k_z T^2 \times \Big[ a_z + 2[(v_x+a_x T)\Omega_y-2[(v_y+a_y T)\Omega_x] \nonumber\\
     & & + \, (z-3v_z T+2x\Omega_y T-2y\Omega_x T)(\Omega_{x}^2+\Omega_{y}^2) \nonumber\\
     & & + \, [(x+3v_x T)\dot{\Omega_y}-(y+3v_y T)\dot{\Omega_x}] \nonumber\\
     & & + \, 2T\Omega_z^2 (x\Omega_y-y\Omega_x)- \Omega_z(x\Omega_x+y\Omega_y) \nonumber\\
     & & + \, xT(\Omega_x\dot{\Omega_z}-\Omega_z\dot{\Omega_x})+yT(\Omega_y\dot{\Omega_z}-\Omega_z\dot{\Omega_y}) \nonumber\\
      & & + \, 3T\Omega_z(v_x\Omega_x+v_y\Omega_y) \Big]
    \label{eq:SingleAIphaseshift}
\end{eqnarray}
where $T$ is the interrogation time.

In the following, we will assume for simplicity that the atom is initially fully centred in the beam, $x=y=0$.
In a gradiometric configuration with two interferometers, denoted A and B, the differential phase shift between the two interferometers, is obtained by direct substraction of Eq.~\ref{eq:SingleAIphaseshift} for both
interferometers:
\begin{eqnarray}
    \Delta\Phi & = &  \kbf.(\gbf^{B}-\gbf^{A})T^2 + 2\kbf.[(\gbf^{B}-\gbf^{A})\times\Ombf]T^3 \nonumber\\
    & & - \, \kbf.(\abf^{B}-\abf^{A})T^2 - 2\kbf.[(\abf^{B}-\abf^{A})\times\Ombf]T^3 \nonumber\\
     & & + \, (z^{B}-z^{A})(\Omega_{x}^2+\Omega_{y}^2)kT^2 \nonumber\\
      & & + \, (v_x^{B}-v_x^{A}) [2\Omega_y+3T\dot{\Omega_y}+3T\Omega_x\Omega_z]kT^2 \nonumber\\
     & & + \, (v_y^{B}-v_y^{A}) [-2\Omega_x-3T\dot{\Omega_x}+3T\Omega_y\Omega_z]kT^2 \nonumber\\
    & & - \, 3(v_z^{B}-v_z^{A})(\Omega_{x}^2+\Omega_{y}^2)kT^3 
      \label{eq:diffPhaseshift}
\end{eqnarray}
where $\gbf^A$ (resp. $\gbf^B$) is the gravity field at the location of interferometer A (resp. B), $\abf^A$ (resp. $\abf^B$) is the acceleration of interferometer A (resp. B) in the Earth's frame, and $z^{A}, v^{A}_{x,y,z}$ (resp. $z^{B}, v^{B}_{x,y,z}$) are the vertical position and velocity coordinates of the atom in interferometer A (resp. B) at the time of the first pulse, expressed in the sensor frame.

\subsection{Requirements based on the differential phase shift Eq.~\ref{eq:diffPhaseshift}}

The first term in Eq.~\ref{eq:diffPhaseshift} is the signal of interest in a gradiometric measurement. Assuming that the gradiometer is perfectly vertical (no tilt), it is given by $\Gamma_{zz} k T^2 b$ with $b$ the gradiometer baseline and $\Gamma_{zz}$ the z-z component of the gravity gradient tensor. The typical gradiometric phase shift associated with the gravity gradient variations that we want our sensor to be able to resolve in this work ($\approx~10E$, commensurate with small gravity anomaly variations on the map) is therefore $\Delta\Phi_\mathrm{crit}=10^{-8} k T^2 b$. In the following, we examine other contributions to the differential phase shift and derive requirements for systematic errors to be bounded to a satisfactory level by imposing that all the other contributions to the phase shift in Eq.~\ref{eq:diffPhaseshift} remain smaller than $\Delta\Phi_\mathrm{crit}$.\\

\paragraph{Static tilts:} In the presence of a static sensor tilt $\alpha$, the gradiometric phase (first term in Eq.~\ref{eq:diffPhaseshift}) becomes $\cos^2 \alpha \Gamma_{zz} k T^2 b$. Imposing that the resulting error, $(1-\cos^2 \alpha)\Gamma_{zz} k T^2 b$, be smaller than $\Delta\Phi_\mathrm{crit}$ sets the requirement $\Gamma_{zz}(1-\cos^2 \alpha)<10E$ which, taking into account that the
Earth's typical gravity gradient is $\Gamma_{zz}\approx 3.10^{-6} s^{-2}$, gives $\alpha<3.3^o$. 
We note that the tilt angle also appears coupled to rotation terms in term 2 of Eq.~\ref{eq:diffPhaseshift}, which is evaluated as $2kT^3\Gamma_{zz} b \cos\alpha\sin\alpha \Omega_y$ assuming the tilt is in the x-direction. This contribution is much smaller than $\Delta\Phi_\mathrm{crit}$ if both the tilt condition $\alpha<3.3^o$ and the below rotation condition Eqs.~\ref{eq:RotReq} are fulfilled.\\

\paragraph{Sensor translational motion:} If the gradiometer undergoes a perfect translational motion (possibly accelerated), no differential phase shift is induced as terms 3 and 4 in Eq.~\ref{eq:diffPhaseshift} involve the \textit{differential} acceleration between the two interferometers.\\

\paragraph{Sensor rotational motion:} Rotations of the gradiometer, in turn, induce different accelerations for the two interferometers, producing non-negligible contributions to terms 3 and 4. By using $z^B-z^A=b$ and the kinematic relation 
\begin{eqnarray}
    \!\!\!\!\! \abf^{(B)}-\abf^{(A)} & \! = \! & \Ombf\times(\Ombf\times\vect{AB})+\dot{\Ombf}\times\vect{AB},
    \label{eq:relativesensormotion2}
\end{eqnarray}
where $\mathbf{AB}$ is the separation vector between the two interferometers A and B [$\mathbf{AB}=(0,0,b)$ in sensor body axes],
terms 3 to 5 in Eq.~\ref{eq:diffPhaseshift} can be evaluated as $2 b k T^2(\Omega_{x}^2+\Omega_{y}^2)-2kbT^3(\Omega_x \dot{\Omega_x}+\Omega_y \dot{\Omega_y})$. We note here that the first term is always positive, meaning that constant rotations induce a positive bias on gravity gradient measurements, as observed for instance on Figs.~\ref{fig:ShipGGmeas}-\ref{fig:FlightGGmeas}. 
A sufficient condition for both these terms to remain bounded below $\Delta\Phi_\mathrm{crit}$ is, for $T=50$ms:
\begin{eqnarray}
    \Omega_{S_{x,y}} & < & 7.10^{-5}s^{-1} \nonumber\\
     \dot{\Omega}_{{S}_{x,y}} & < & 7.10^{-4}s^{-2} 
     \label{eq:RotReq}
\end{eqnarray}

\paragraph{Atomic velocities and temperature:} Terms 6 to 8 in Eq.~\ref{eq:diffPhaseshift} involve, for each interferometer, the initial velocity of the atom at the time of the first pulse. Assuming a typical thermal velocity, $\sqrt{k_B\theta/m}$, the requirement that these terms remain smaller than $\Delta\Phi_\mathrm{crit}$ yields a stringent constraint on transverse temperature, which critically depends on the sensor rotation rate and rotational acceleration; as an example, for $b=1m$, assuming $\Omega_S=7\times10^{-6}s^{-1}$ and $\dot{\Omega_S}=7\times 10^{-5}s^{-2}$ (so that we are well into a regime where Eq.~\ref{eq:RotReq} is fulfilled and purely rotational effects are negligible), we would find $\theta_{x,y} < 5$nK~\footnote{The constraint on z-temperature is not stringent here, $\theta_z <  1.8K$ for rotations rates given by Eq.~\ref{eq:RotReq}.}. In situations where this condition breaks, such as in the "moderate" scenario considered in this work, the interplay between platform rotation and atomic thermal velocities is therefore expected to contribute a phase shift. An exact calculation would however require a proper treatment of the thermal cloud of atoms, as done in this work, going beyond the simplified derivation presented in this Appendix.\\

\subsection{Further considerations}
\label{sec:AppSubsec2}

\paragraph{Effects of gravity gradients:} The above calculation has neglected variations of the gravity field \textit{inside each individual interferometer}. At lowest order in the gravity gradient, the phase shift of an interferometer in the presence of a gravity gradient is given by
$\kbf.\gbf T^2+\kbf.(\rbf_1+\vbf_1 T+7\mathbf{a}T^2/12) T^2\Gamma_{zz}$~\cite{Peters2001}. When considering a differential measurement between two interferometers A and B, the first two terms produce the gradiometric signal of interest $\Gamma_{zz}kT^2b$, while the two others give rise, at lowest order in the gravity gradient, to corrections to the differential phase shift
$$\kbf.(\vbf_1^B-\vbf_1^A) T^3\Gamma_{zz}-\dfrac{7}{12}\kbf.(\abf^{(B)}-\abf^{(A)})T^4\Gamma_{zz}.$$
Assuming thermal velocities for $\vbf_1^A$ and $\vbf_1^B$, and for $T=50$ms, we find that the first term will remain smaller than $\Delta\Phi_{\mathrm{crit}}$ provided the z-temperature is below $50\mu K$.
In turn, the requirement that the second term is smaller than $\Delta\Phi_{\mathrm{crit}}$ sets a constraint on the sensor rotation rate (by virtue of Eq.~\ref{eq:relativesensormotion2}), $\Omega_{S_{x,y}} < 1s^{-1}$, which is however much less stringent than Eq.~\ref{eq:RotReq}.
An more exact treatment of gravity gradient effects capturing higher couplings with sensor motion and tilts would require solving Eq.~\ref{eq:ExtEOM} in the presence of a inhomogeneous gravity field. While this goes beyond the scope of this Appendix, this is captured in our numerical interferometer model.\\

\paragraph{Effects of pulse infidelities:} 
All the above requirements are based solely on a simplified, single-atom, phase shift calculation.
As such, it cannot capture cross-couplings between dynamic motion and inhomogeneous atomic clouds.
Additionally, even in the single-atom case, there are other physical processes at play that this approach overlooks. This notably includes effects arising from pulse infidelities due to the motion of atoms in the beam.
An important one is the effect of platform accelerations. An acceleration of the platform does not cause any differential phase shift as it is common to both interferometers, but still causes, in each separate interferometer, pulse infidelity leading to a loss of fringe contrast and an apparent phase shift. This includes both longitudinal accelerations, which induce additional Doppler shifts, and transverse accelerations, which cause the atoms to drift off-centred in the beam under the action of the centripetal force. 
As a rough estimate, longitudinal accelerations will typically start to play a role when the Doppler shift at the final pulse, $2Ta_z k_{\mathrm{eff}}$, is of the order of the pulse bandwidth $\Omega_R$, which gives $a_z\approx0.5ms^{-2}$ for the "moderate" sensor specifications and $a_z\approx3ms^{-2}$ for the "enhanced" sensor specifications.
As for transverse accelerations, a simple condition is that the atoms should remain within the beam during the full interrogation sequence, $a_{x,y}(2T)^2/2<<r_\mathrm{beam}$ with $a_{x,y}$ the transverse acceleration of the considered interferometer, which typically gives $a_{x,y}<< 0.2 ms^{-2}$ for $T=50$ms and a 1-mm beam radius. Importantly though here, $a_{x,y,z}$ would depend not only on the platform acceleration, but also on the platform rotation rate and the distance between the considered interferometer and the platform centre of rotation, through a relation similar to Eq.~\ref{eq:relativesensormotion2}. Sensor tilts, which accentuate the atoms getting off-centered as they free-fall in the beam, and other inertial forces, which non-trivially affect atomic trajectories, can make these criteria significantly more involved - not to mention further complexity arising from atomic cloud inhomogeneities which lead to atom-dependent trajectories. An exact treatment requires as accurate a model as the one presented in this work, to account for the interplay between non-inertial atomic trajectories and atom-light interactions. 
\\

In summary, going beyond the calculation presented in this Appendix, our atom interferometer model allows for an exact treatment: (i) not relying on an approximate solution for the atomic external dynamics; (ii) accounting for variations in the experienced external fields, e.g. $\gbf$, during the interferometric sequence; (iii) capturing couplings between the ballistic motion and the internal dynamics of atoms at the level of each atom-light interaction - and not only at the phase shift level; (iv) capturing inter-atom dephasings and cross-couplings between the inhomogeneous cloud, inertial forces and light pulses; (v) accounting for variations in inertial fields between ellipse shots, rather than relying on a phase shift obtained at fixed external fields.


\section{Map-matching using a particle filter algorithm}
\label{sec:AppPF}

\subsection{Description}

In a particle filter, the position estimate probability density function (PDF) is represented by a set of particles, each of which is weighted according to the probability that the particle is at the true position. The particle filter is a Bayesian estimator in the sense that between measurement intervals, the particles are propagated to account for the new position of the platform and the uncertainty in the change of position. During the measurement stage, the innovation is calculated for each particle and converted to a weighting using the measurement likelihood function. Periodically, particles are also resampled, whereby particles in ‘low probability’ regions are deleted and new particles added to ‘high probability’ regions. 

The typical implementation follows a four-step process:
\paragraph*{Initialisation:} Particles are randomly distributed to form the initial PDF - this would be set according to the latitude / longitude uncertainty ellipse as estimated by the navigation filter. Each particle has an equal weight, and the particles are distributed according to a bivariate Gaussian distribution.
\paragraph*{Propagation:} As the platform is in motion, the particles must move with it at each timestep. In the propagation phase, the entire particle net at the previous timestep is first translated according to the navigation filter's estimate of the position change between the previous and current timestep. In addition, each individual particle has a translation according to the velocity uncertainty. The effect is that the particles both shift collectively and spread out to capture the fact that the position change estimate is uncertain due to velocity errors. Again, all particle weightings remain equal. The PDF at this stage represent the 'prior' estimate. 
\paragraph*{Measurement:} The sensor measurement is now used to inform the weighting of each particle. In this application, the measured gravity gradient would be compared to the gravity gradient at each particle location on the reference map to form an \textit{innovation}, which is then converted to a weighting via a measurement likelihood function. Particles that have a close match would receive a high weight, and particles that have a poor match would receive a low weight. The PDF at this stage now represents the 'posterior' estimate.
\paragraph*{Resampling:} The particles are now redistributed whereby particles with low weight are eliminated, and particles added around those with high weight, such that the total number of particles is maintained. The particle weighting is now reset such that all particles have the same weight. The state estimate at this timestep may be made by taking the mean and covariance of the particle net.

\subsection{Implementation details}

The general algorithm tuning characteristics that we have adopted all throughout this work are as follows:\\
\qquad - 5000 particles are maintained at all times.\\
\qquad -  During the measurement phase, the particles are weighted by sampling a measurement PDF defined as a Gaussian distribution with zero mean and standard deviation equal to $2\sigma$ with $\sigma$ the gradiometry measurement noise.\\
\qquad -  The propagation and measurement phases run at 1Hz, or the gradiometry sample rate, whichever is slower.\\
\qquad -  The particles are resampled and an estimate provided every 40x measurements for flight trajectories (resp. 400x for maritime trajectories), resulting in an update rate of 40s (resp. 400s) for a 1Hz measurement rate. The resampling scheme used is a cumulative density function.\\
\qquad -  Following resampling, the particles are redistributed using a bivariate distribution maintaining the same covariance. This is to avoid particle clustering / multimodal distributions.\\


\end{document}